\documentclass[letterpaper,11pt]{article}

\usepackage{dynkin-diagrams}
\usepackage{cite}
\usepackage{amsmath}
\usepackage{amssymb}
\usepackage{amsthm}
\usepackage{mathrsfs}
\usepackage{xypic}
\usepackage{tikz-cd}

\usepackage{hyperref}
\usepackage{color}
\definecolor{darkred}{rgb}{0.65,0.15,0}
\hypersetup{pdfborder={0 0 0},colorlinks=true,urlcolor=darkred,citecolor=blue,linkcolor=darkred,linktocpage=true}

\usepackage{float} 

\usepackage{geometry}

\newgeometry{vmargin={35mm}, hmargin={35mm}} 

\usepackage{keyval}
\makeatletter
\define@key{setpar}{left}[0pt]{\leftmargin=#1}
\define@key{setpar}{right}[0pt]{\rightmargin=#1}
\define@key{setpar}{both}{\leftmargin=#1\relax\rightmargin=#1}
\makeatother

\makeatletter
\renewcommand*\env@matrix[1][\arraystretch]{%
  \edef\arraystretch{#1}%
  \hskip -\arraycolsep
  \let\@ifnextchar\new@ifnextchar
  \array{*\c@MaxMatrixCols c}}
\makeatother  

\def\eg{{\it e.g.}}
\def\ie{{\it i.e.}}

\def\ad#1{\mathrm{ad}(#1)}
\def\End{\mathrm{End}}

\def\DWeight#1#2#3{\bigl(\raise2.5pt\hbox{${}_{#1}$}{}^{#2}_{#3}\bigr)}

\def\AAWeight#1#2{\bigl(\raise0pt\hbox{${}^{#1}_{#2}$}\bigr)}

\def\fg{{\mathfrak g}}

\def\fk{\mathfrak{k}}
\def\fp{{\mathfrak p}}

\def\nn{\nonumber}
\def\fk{\mathfrak{k}}

\def\*{\partial}

\def\transpose{\intercal}

\def\Red#1{\textcolor{red}{#1}}

\def\RR{{\mathbb R}}
\def\ZZ{{\mathbb Z}}

\def\LL{{\mathscr L}}

\newcommand{\dd}{{\mathsf{d}}}
\newcommand{\KK}{{\mathsf{K}}}

\def\bra#1{\langle#1|}
\def\ket#1{|#1\rangle}
\def\brabra#1{\langle\hspace{-0.8mm} \langle#1|\hspace{-0.4mm} |}
\def\ketket#1{|\hspace{-0.4mm} |#1\rangle\hspace{-0.8mm}\rangle}

\def\sh{\sharp}
\def\fl{\flat}

\def\cn#1{{C_{#1}}}

\def\shift#1#2{\underset
  {\scriptscriptstyle\Red{#1}}{#2{}^{\mathstrut}_{\mathstrut}}}

\numberwithin{equation}{section}
\numberwithin{figure}{section}
\numberwithin{table}{section}

\newlength\symlength
\newcommand\lplus{
 \ensuremath{
  \mathop{
   \begin{tikzpicture}[line width=0.12ex]
    \useasboundingbox (-1ex, -0.7ex) rectangle (.7ex, 1ex);
\draw (60:\symlength) arc (60:300:\symlength);
\draw (-.75\symlength,0) -- (.75\symlength,0);
\draw (0,-.75\symlength) -- (0,.75\symlength);
   \end{tikzpicture}
  }\nolimits
 }
}

\def\inplus{\lplus}

\def\gG{\rm G}
\def\gP{\rm P}
\def\gK{\rm K}

\begin{document}

\frenchspacing

\includegraphics[height=1cm]{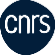}
\hspace{2mm}
\includegraphics[height=1.1cm]{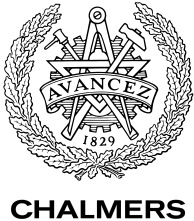}
\hspace{2mm}
\includegraphics[height=1cm]{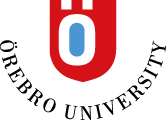}

\vspace{-12mm}
{\flushright Gothenburg preprint, CPHT-RR002.022024 \\ 
  %\today\\
}
%March, 2021\\} %July, 2019\\}

\vspace{4mm}

\hrule

\vspace{16mm}

%{\flushright {\tt \today}\\
%{\tt linfty4.tex}\\
%}
%\null\vspace{1cm}

\thispagestyle{empty}

\begin{center}
  {\Large \bf \sc Teleparallel Geroch geometry}
    \\[14mm]

{\large
Guillaume Bossard${}^1$, Martin Cederwall${}^2$ and Jakob Palmkvist${}^3$}

\vspace{14mm}
       {\footnotesize ${}^1${\it Centre de Physique Th\'eorique, CNRS, Institut Polytechnique de Paris,\\
91128 Palaiseau cedex, France}}

\vspace{2mm}
       {\footnotesize ${}^2${\it Department of Physics,
         Chalmers Univ. of Technology,\\
 SE-412 96 Gothenburg, Sweden}}

\vspace{2mm}
       {\footnotesize ${}^3${\it Department of Mathematics,
         \"Orebro Univ.,\\
 SE-701 82 \"Orebro, Sweden}}

\end{center}

\vfill

\begin{quote}
\textbf{Abstract:} 
We construct the teleparallel dynamics for extended geometry where the structure algebra is (an extension of) an untwisted affine Kac--Moody algebra. This provides a geometrisation of the Geroch symmetry appearing on dimensional reduction of a gravitational theory to two dimensions. The formalism is adapted to the underlying tensor hierarchy algebra, and will serve as a stepping stone towards the geometrisation of other infinite-dimensional, \eg\ hyperbolic, symmetries.
\end{quote} 

\vfill

\hrule

\noindent{\tiny email:
  guillaume.bossard@polytechnique.edu, martin.cederwall@chalmers.se, jakob.palmkvist@oru.se}

\newpage

\tableofcontents

\newpage

\section{Introduction}

Extended geometry \cite{Palmkvist:2015dea,Cederwall:2017fjm,Cederwall:2018aab,Cederwall:2019bai} is a general framework for geometrising duality symmetries in gravitational theories, thus in a certain sense incorporating them in the full dynamics.
Special cases are provided by exceptional geometry
\cite{Hull:2007zu,Pacheco:2008ps,Hillmann:2009pp,Berman:2010is,Berman:2011pe,Coimbra:2011ky,Coimbra:2012af,Berman:2012vc,Park:2013gaj,Cederwall:2013naa,Cederwall:2013oaa,Aldazabal:2013mya,Hohm:2013pua,Blair:2013gqa,Abzalov:2015ega,Hohm:2013vpa,Hohm:2013uia,Hohm:2014fxa,Cederwall:2015ica,Bossard:2017wxl,Bossard:2017aae,Bossard:2018utw,Bossard:2019ksx,Bossard:2021jix,Bossard:2021ebg,Bossard:2023ajq}.

Recently, it has been realised that a teleparallel formulation  
\cite{Cederwall:2021xqi,Cederwall:2023xbj}
provides a version of extended geometry that has several advantages. It is arguably more geometric than the standard coset dynamics, and the connection to tensor hierarchy algebras % (THA's)
\cite{Palmkvist:2013vya,Bossard:2017wxl,Carbone:2018xqq,Cederwall:2019qnw,Cederwall:2021ymp,Cederwall:2022oyb,Cederwall:2023stz} gives a possibility to construct \cite{Cederwall:2023xbj} a full Batalin--Vilkovisky action \cite{Batalin:1981jr}. In addition, since the module of the embedding tensor 
\cite{deWit:2008ta}
arises naturally (as torsion), the formalism is the most natural one for obtaining gauged supergravities.

When gravity, or theories containing gravity, are dimensionally reduced to two dimensions, an affine symmetry appears as a global symmetry, relating solutions to each other. This is the so called Geroch symmetry \cite{Geroch:1972yt}. It is the untwisted affine extension of the Ehlers symmetry \cite{Ehlers:1957zz}, which for pure $d$-dimensional gravity is $A_{d-3}$. In the presence of other massless fields it can be extended; in string theory/M-theory the Ehlers symmetry is enhanced to $E_8$ and the Geroch symmetry to $E_9$
\cite{Julia:1981wc,Julia:1982gx,Nicolai:1987kz}.
Affine extended geometry---geometrisation of Geroch symmetry---has been constructed earlier \cite{Bossard:2017aae,Bossard:2018utw,Bossard:2021jix}, and applied to gauged supergravity \cite{Bossard:2022wvi,Bossard:2023wgg,Bossard:2023jid}.
The present work provides a teleparallel reformulation.
Apart from the already mentioned possible applications,
we view it mainly as a stepping stone towards other cases with infinite-dimensional structure groups, in particular 
the geometrisation of Belinskii--Khalatnikov--Lifshitz (BKL) symmetry \cite{BKL,Damour:2001sa,Damour:2002cu,Henneaux:2007ej}
as extended geometry with (an extension of) an over-extended Kac--Moody algebra as structure algebra. We comment more on this issue in Section \ref{OutlookSection}.

The purpose of this paper is thus to construct and present the teleparallel version of extended geometry with (a slight extension of) an untwisted affine algebra as structure algebra. The use of the teleparallel formalism avoids the difficulties associated with eliminating the spin connection by setting torsion to zero, and is more geometric than the coset dynamics (where generalised diffeomorphisms are not manifest). In addition, it naturally connects to the tensor hierarchy algebras, where the full Batalin--Vilkovisky fields of extended geometry are identified as elements in a grading. We expect this to be even more useful as a guideline when addressing larger structure algebras, and believe that some of the lessons learned presently will facilitate that pursuit.
Finally, the teleparallel Lagrangian is especially well suited for generalised Scherk--Schwarz reduction \cite{Hohm:2014qga}. In particular, the form of the supergravity potential derived in \cite{Bossard:2023wgg} follows directly in the teleparallel formulation.

There are a few differences in conventions between the present paper and earlier work on affine extended geometry
\cite{Bossard:2017aae,Bossard:2018utw,Bossard:2021jix}.
The main one is that the extension of the affine algebra (described in Section \ref{NotationSection}) contains the Virasoro generator $L_1$ instead of $L_{-1}$, and that the r\^oles of highest and lowest weight modules are interchanged.
The reason for our choice is  that we want to adapt to established conventions for tensor hierarchy algebras (agreeing with conventions for contragredient superalgebras), where grading with respect to some node of a Dynkin diagram produces lowest weight modules at  
positive degrees. In this sense, our conventions agree with and extend those of refs.
\cite{Cederwall:2019qnw,Cederwall:2019bai,Cederwall:2021xqi,Cederwall:2023xbj}.

\section{Affine algebras and representations}

\subsection{Algebra and notation\label{NotationSection}}

An untwisted affine Kac-Moody algebra $\fg^+$ is a centrally extended loop algebra spanned by generators
$T_{A,m}$ and $\KK$. The index $A$ labels a finite-dimensional semi-simple Lie algebra $\fg$, $m\in\ZZ$ is a ``mode number'', and $\KK$ a central generator. The non-vanishing Lie brackets are
\begin{align}
[T_{A,m},T_{B,n}]=f_{AB}{}^CT_{C,m+n}+m\eta_{AB}\delta_{m+n,0}\KK\;.
\end{align}
Here, $f_{AB}{}^C$ are structure constants of $\fg$ and $\eta_{AB}$ its Cartan--Killing metric.
We will always consider the split real form. Generalisation to other real forms is straightforward
\cite{Bossard:2023ajq}.

The Sugawara construction provides Virasoro generators acting on modules with eigenvalue $k$ of the central generator $\KK$ according to
\begin{align}
L_m^{(k)}={1\over2(k+g^\vee)}\sum_{n\in\ZZ}\eta^{AB}:T_{A,n}T_{B,m-n}:\;,
\end{align}
where the $T$'s are the respective representation matrices and $g^\vee$ is the dual Coxeter number of $\fg$. They act on $\fg^+$ as
$[L_m,T_{A,n}]=-nT_{A,m+n}$.
The central charge is 
$c^{(k)}={k\dim\fg\over k+g^\vee}$. 

The ``structure algebra'' we will be working with is the semidirect sum $\fp\inplus\fg^+$, where $\fp$ is the parabolic subalgebra of the Virasoro algebra spanned by $L_0$ and $L_1$. %{\Red{$\dd$ is not defined.}}
%We define the derivation $\dd$ of the affine algebra, that acts on a module of weight $w$ as $\dd = \dd + w $. 
Generators of $\fp\inplus\fg^+$ are labelled by indices $\alpha,\beta,\ldots$. 
We use the values $0$ and $1$ of the index for $L_0$ and $L_1$, respectively.
(Occasionally, we will use indices $\hat\alpha,\hat\beta,\ldots$ to label the extension of $\fg^+$ with all Virasoro generators.)
%$\hbox{Vir}\inplus\fg^+$.
The additional non-vanishing brackets in $\fp\inplus\fg^+$ thus are
\begin{align}
[L_0,T_{A,m}]&=-mT_{A,m}\;,\nn\\
[L_1,T_{A,m}]&=-mT_{A,m+1}\;,\\
[L_0,L_1]&=-L_1\;.
\end{align}

We will often refer to mode-shifted generators. Define a shift operator $s$ by
\begin{align}
sT_{A,m}&=T_{A,m+1}\;,\nn\\
s\KK&=0\;,\\
sL_m&=L_{m+1}\;,\nn
\end{align}
and define $T^{(n)}_\alpha=s^nT_\alpha$. 
We also allow $n<0$ by letting $s^{-1}$ have the same kernel as $s$.
Note that this produces results outside $\fp\inplus\fg^+$.
%(but in $\hbox{Vir}\inplus\fg^+$).
Mode-shift is not an automorphism. In particular, it annihilates $\KK$. When mode-shifted generators are used later, \eg\ in the construction of torsion, this has to be compensated.

Generalised vectors are assigned to the lowest weight module $R(-\lambda)$ at level $k=1$.
%{\bf GB: I set $k=1$, I do not think it makes sense to have negative level in the formula above for the Sugawara construction}. 
$\lambda$ is the fundamental weight dual to the affine root, and is light-like, $(\lambda,\lambda)=0$. We label elements in this module by an upper index $M,N,\ldots$, \ie, $V^M$. Most expressions and calculations however use an index-free notation, where elements in this lowest weight fundamental module are treated as bra vectors,
$\bra V$, and elements in the dual highest weight module $R(\lambda)$ are ket vectors, $\ket W$.
Representation matrices in this fundamental representation are written in lowercase as $t_\alpha$, and $\ell_m$ is used for the representation matrices of Virasoro generators.

Although it is useful to introduce the whole Virasoro algebra, the theory is only invariant under the Lie algebra  $\fp\inplus\fg^+$. Because this Lie algebra is unchanged under the automorphism defined by 
\begin{align}
L_0&\mapsto L_0-w\KK\;,\nn\\
L_1&\mapsto L_1\;,\nn\\
T_{A,m}&\mapsto T_{A,m}\;\\
\KK&\mapsto\KK\; , \nn
\end{align}
one can define its modules with an arbitrary {\it weight} $w$. Then one can define the affine extension derivation $\dd$ as acting on the basic module of weight $w$ as $\dd \ket W = ( \ell_0 - w) \ket W$. Irrespective of the weight, the highest weight state
$\ket0$ obeys $\ell_0\ket0=0$.
The standard weight we will use for highest weight fundamentals (``covectors'') is $w=1$, and for lowest weight fundamentals
(``vectors'') $w=-1$. The reason is tensorial transformation under generalised diffeomorphisms, see
Section \ref{GenDiffSection}. 

%Fundamental highest and lowest weight modules. Shifted modules.

\subsection{Involution and compact subalgebra}

A generalised vielbein $E$ parametrises the coset $(\gG^+\rtimes \gP)/\gK(\gG^+)$, where $\gP$ is the parabolic subgroup of 
$\mathrm{SL(2)}$ generated by $L_0$ and $L_1$.
We parametrise it as $E=e^{-\phi L_1}\varrho^{-\dd} e$, where $e\in \gG^+$.
We will use the vielbein in the fundamental representation with $w=1$. It should, as usual, be seen as a matrix
$E_M{}^A$, where $A$ is a ``flat'' index.
It thus becomes
%We choose a parametrisation of the vielbein $E$:
%\footnote{One may of course choose the opposite order  $e^{\phi' L_1}\varrho^{-L_0}$. Then $\phi'={\phi\over\varrho}$. {\Red{I would remove the footnote, because obvious.}}}
\begin{align}
E=e^{-\phi\ell_1}\varrho^{-\ell_0+1} e\;,\label{EDef}
\end{align}
The additional $\varrho$ factor is present since we construct $E$ as the group element in the
$w=1$ representation. Equivalently, it is introduced to ensure that $E_M{}^A$ transforms as a covector under generalised diffeomorphisms, and not as a density (see Section \ref{GenDiffSection}). The metric $G_{MN}$ would  formally be given as 
\begin{align}
G=EHE^\transpose=\varrho^{2}e^{-\phi \ell_1}\varrho^{-\ell_0} g\varrho^{-\ell_0^\transpose}e^{-\phi\ell_1^\transpose}\;,
\label{MetricParam}
\end{align}
with $g=eHe^\transpose$ being the $\gG^+$ metric. $H$ is the metric corresponding to the Chevalley involution
\begin{align}
\tau_H(T)=-HT^\transpose H^{-1}\;, \label{Hinvolution}
\end{align}
such that the Lie algebra $\mathfrak{k} \subset \mathfrak{g}^+$ defining $\gK(\gG^+)$ is invariant, i.e. $\tau_H$ has the eigenvalue $1$ on $\mathfrak{k}$. 
%\Red{Decide on order of $\dd$ and $L_1$. Maybe inconsistent now.}

For finite-dimensional Lie algebras and representations, $G$ is a well-defined group element.
However, since we now are dealing with an infinite-dimensional algebra, and its infinite-dimensional highest or lowest weight representations, we must be more precise about the definition of the group and its modules. The group element $e$ parametrises the fields of the theory and is defined in the maximal positive Borel extension of the Kac--Moody group \cite{Marquis}.  This means that the generalised vielbein $E$ includes elements in the universal enveloping algebra of arbitrary high $L_0$ degree, as for example in $e^{-\phi L_1}$. When we write $(\gG^+\rtimes \gP)$, we always mean the maximal positive Borel extended group. 
The positive extended group $(\gG^+\rtimes \gP)$ acts consistently on elements of the highest weight module $R(\lambda)$ that have a maximal $L_0$ degree. Therefore vectors $|W\rangle\in R(\lambda)$ are always understood to only carry finitely many non-zero components in a chosen basis. This is justified in the physical model because the objects in this modules are derivatives of the fields that satisfy the section constraint \cite{Bossard:2018utw}.  On the contrary $(\gG^+\rtimes \gP)$ acts consistently on elements of the completed lowest weight module $R(-\lambda)$, that include formal vectors with unbounded $L_0$ degree and infinitely many non-vanishing components in a given basis.\footnote{This is again justified in the physical model because objects in $R(-\lambda)$ are for example the two-dimensional vector fields in exceptional field theory.} These modules are naturally dual to each other since the scalar product $\langle V|W\rangle$ is always finite. We refer to \cite{Marquis,Peterson:1983,Carbone:2003} for the proper definition of the maximal positive Borel completed group and its modules. This definition does not allow to act with the generalised metric $G$ on $R(\lambda)$, 
and therefore we do not want to define $G$ as a group element. However, $G^{-1}$  is a well defined bilinear form on $R(\lambda)$. To understand this, let us define $H = h h^\intercal $ for an element $h$ of the minimal Kac--Moody group $\gG^+_0$ generated by finite products of real roots generators \cite{Peterson:1983}. It is necessary to define $h\in \gG^+_0$ because $\gK(\gG^+) \subset \gG^+_0 \subset \gG^+$, such as to act equivalently on highest and lowest weight modules.   There is a preferred basis $\langle e_A|$ in which $ \langle e_A | H =   \langle e_A | $ and one can define the inverse generalised metric as an element of $R(-\lambda)\otimes R(-\lambda)$ in this preferred basis as
\begin{align} 
\brabra {G^{-1}}  = \brabra H  ( E^{-1} \otimes E^{-1} ) =  \delta^{AB} \langle e_A | E^{-1} \otimes  \langle e_B | E^{-1} \; . 
\label{GInverseDef}
\end{align}
Note that the equivalent definition for $G$ does not belong to the module $R(\lambda)\otimes R(\lambda)$. In components one may write  the scalar product $G^{MN} V_M W_N$ that is well defined, whereas  $G_{MN} V^M W^N$ does not make sense with the definitions above. 
Note also, that the 
%(somewhat awkward) 
definition \eqref{GInverseDef} coincides with the usual definition
for the metric in finite-dimensional modules of finite-dimensional algebras in components with $G^{MN}=\delta^{AB} E_A{}^M E_B{}^N$. In practice one may use the formal expressions, as long as one understands that all matrix multiplications must be done in an order in which all expressions make sense at every step. 

%
%
%On the contrary, $H$ is defined in the small Kac--Moody group, and only includes elements of $\dd$-bounded degree in the universal enveloping algebra. There is no well defined notion of a group that is extended both positively and negatively in $\dd$ degree, and $G$ is therefore not an element in a well defined group. The positive Borel extended group that we write 
%
%
%there are issues with well-defined\-ness of the metric. In particular, it may not be well defined as a group element, and scalar products of the type $G_{MN}V^MW^N$ may lead to infinite expressions.
%We should therefore avoid expressions with the metric,
% unless we can assure that they are well defined. One instance that does not present any risk is expressions like
% $G^{MN}V_MW_N$, where $V$ and $W$ are elements of finite depth in a highest weight module, \ie, annihilated by applying some finite number of raising operators.
% The metric $H_{AB}$ and its inverse are always well defined. In a weight basis, they can be taken as the unit matrix.

%\Red{At some point, it might be meaningful to define the  vielbein in the fundamental 
%as $E=e^{\phi\ell_1}\varrho^{\ell_0+1} e$, so that it carries weight 1, which is the canonical (tensorial) weight for a covector, so that contractions with vectors give scalars.}

It will be convenient to define the Chevalley involution \eqref{Hinvolution}  conjugated by $E$ as
\begin{align}
\tau(T)=E\tau_H(E^{-1}TE)E^{-1}\;.\label{ChevInvDet}
\end{align}
Note that this operation is an involution on %the maximal positive Borel completed algebra 
$\fg^+$, but not on $\fp\lplus\fg^+$.  The involution singles out the conjugate of the  compact subalgebra $\fk$ on which $\tau_H$ has the eigenvalue $1$. The invariant algebra under $\tau$ is a subalgebra of $\fg^+$, but does not act equivalently on $R(\lambda)$ and $R(-\lambda)$. This involution extends to the full (completed)  algebra where $\fg^+$ is extended by all Virasoro generators, but it does not preserve the parabolic subalgebra $\fp\subset\hbox{Vir}$. 
 The r\^ole of $\fp$ is to ``twist'' the involution on $\fg^+$.
 Note that eq. \eqref{ChevInvDet} amounts to ``flattening'' the indices of a matrix $T_M{}^N$, then applying the Chevalley involution, and finally converting back to ``curved'' indices. It thus formally coincides with the usual involution 
 $\tau(T)=-GT^\transpose G^{-1}$ for finite-dimensional algebras and modules.

%{\it old version: We are interested in the involution %on the Lie algebra $\fp\inplus\fg^+$ 
%implemented by the vielbein $E$.
%For finite-dimensional groups, the involution is defined in terms of conjugation by $G$ as
%\begin{align}
%\tau(T)=-GT^\transpose G^{-1}\;.\label{GInvolution}
%\end{align}
%For the Kac--Moody group we defined equivalently the Chevalley involution using \eqref{Hinvolution} such that eq. 
%\eqref{GInvolution} is replaced by 
%\begin{align}
%\tau(T)=E\tau_H(E^{-1}TE)E^{-1}\;.
%\end{align}
%Note that this operation is an involution on $\fg^+$, but not on $\fp\lplus\fg^+$. If one extends to the full algebra $\hbox{Vir}
%\lplus\fg^+$, it is an involution, but it does not preserve the parabolic subalgebra $\fp\subset\hbox{Vir}$. 
%This is not a problem. The involution is used to single out a compact subalgebra $\fk(\fg^+)$, on which $\tau$ has the eigenvalue 
%$1$. It is a subalgebra of $\fg^+$. The r\^ole of $\fp$ is thus to ``twist'' the involution on $\fg^+$.}

%\Red{Compact subalgebra entirely in $\fg^+$ etc... What happens with generators in $\fp$...}

Keeping the explicit parametrisation \eqref{MetricParam} in terms of $L_0$ and $L_1$, one readily arrives at
\begin{align}
\tau(T)=e^{-\phi\,\ad{L_1}}\varrho^{-\ad{L_0}}\tau_e(\varrho^{\ad{L_0}}e^{\phi\,\ad{L_1}}T)\;,
\label{TauG}
\end{align}
where $\tau_e(T)=e\tau_H(e^{-1}Te)e^{-1}$.

The involution $\tau_e$ can be seen as a $\fg$ involution, varying over the circle. When $e$ is the unit element, it simply implies 
$\tau_1(T_{A,m})=T_{A^\star,-m}$, where $A^\star$ expresses the Chevalley involution of $\fg$, \ie, it becomes the Chevalley involution on $\fg^+$.

We will have occasion to perform the involution on mode-shifted generators $T^{(n)}$, in particular for $n=-1$.
It is obvious from the above (since $\tau_e$ acts as a $\fg$ involution locally on the circle) that 
\begin{align}
\tau_e(T^{(n)})=(\tau_e(T))^{(-n)}\;.
\label{ModeInvolutionOne}
\end{align}
%\Red{Write with $s$.}
This will change with the introduction of $L_0$ and $L_1$, which are diffeomorphisms.

The easiest way to understand the impact of these diffeomorphisms on the involution $\tau$ is to consider $L_0$ and $L_1$ as scaling and translation in a variable ${1\over z}$:
\begin{align}
L_0=-z{d\over dz}\;,\quad L_1=-z^2{d\over dz}\;.
\end{align}
Then, 
\begin{align}
\varrho^{L_0}\;:\quad&{1\over z}\mapsto{\varrho\over z}\;,\nn\\
e^{\phi L_1}\;:\quad&{1\over z}\mapsto {1\over z}+\phi\;.\label{PonZAction}
\end{align}
Mode number is identified with degree of homogeneity in $z$ Mode-shift by $n$ units is identified with multiplication by $z^n$, and involution $\tilde\tau$ as $z\mapsto {1\over z}$.
Using the decomposition of $\tau$ of eq. \eqref{TauG},
\begin{align}
{1\over z}\mapsto ( e^{-\phi L_1}\circ\varrho^{-L_0}\circ\tilde\tau\circ\varrho^{L_0}\circ e^{\phi L_1}){1\over z}
=\phi+\varrho^2{z\over1-\phi z}\equiv f(z)\;.
\end{align}

This gives the modification of eq. \eqref{ModeInvolutionOne}:
\begin{align}
\tau(T^{(-n)})=(f(s))^n\tau(T)\;
\label{ModeInvolution}
\end{align}
where $s$ is the shift operator, $T^+=sT$.
In particular,
\begin{align}
\tau(T^-)=f(s)\tau(T)=\phi\tau(T)
+\varrho^2\sum_{n=0}^\infty\phi^n(\tau(T))^{(n+1)}\;.
\label{TauTMinus}
\end{align}
Keeping expressions of this type under control will be instrumental for demonstrating local $\gK(\gG^+)$ invariance of the action in Section
\ref{ActionSection}. %\Red{(Maybe not anymore?)}

%\Red{Something on compact sub-algebra and representations, refs. ... but we do not use...}
Involutory subalgebras of infinite-dimensional Kac--Moody algebras are more complicated than for the finite-dimensional ones
\cite{Nicolai:2004nv,Damour:2006xu,Kleinschmidt:2006dy,Kleinschmidt:2018hdr,Kleinschmidt:2021agj}.
In particular, they are not semi-simple, and the presence of non-trivial ideals allow the existence of \eg\ finite-dimensional ``spinor'' representations. Little is known even about the behaviour of highest weight modules of affine Kac--Moody algebras under their compact subalgebra. Such knowledge would be ideally suited for the present project. We will instead only assure that the dynamics we formulate is invariant under local transformations in the compact subalgebra.

%In the following the subscript $G$ will be dropped and the involution referred to as $\tau$.

\subsection{Invariant tensors and identities\label{InvariantTensorsSection}}

We use the notation $\vee$ and $\wedge$ for symmetric and antisymmetric tensor product, both for elements in fundamentals (``states'') and operators on the tensor product, with the normalisation $a\otimes b=a\vee b+a\wedge b$.
The permutation operator on states is denoted $\varsigma$: $\varsigma(\ket U\otimes \ket V)=\ket V\otimes \ket U$.

We will need some properties of tensor products of fundamentals.
Consider a product of highest weight fundamentals, $\ketket W=\ket U\otimes\ket V$.
Having $k=2$, they are naturally acted on by tensor product,
$T_\alpha\cdot\ketket W=(1\otimes t_\alpha+t_\alpha\otimes1)\ketket W$.
Since both $L^{(2)}_m$ and $1\otimes\ell_m+\ell_m\otimes1$ transform the generators in $\fg^+$ the same way,
the differences 
\begin{align}
L^{\textrm{coset}}_m=1\otimes\ell_m+\ell_m\otimes1-L^{(2)}_m\;,
\end{align}
which also generate a Virasoro algebra (with central charge 
$2c^{(1)}-c^{(2)}={2\dim\fg\over(1+g^\vee)(2+g^\vee)}$), are invariant under $\fg^+$. 
These coset generators thus provide invariant tensors, although not under the whole $\fp\inplus\fg^+$.
We prefer to rescale them and use
\begin{align}
C_m=(2+g^\vee)L^{\textrm{coset}}_m=1\otimes\ell_m+\ell_m\otimes1-\sum_{n\in\ZZ}\eta^{AB}t_{A,n}\otimes t_{B,m-n}\;.
\label{CmEq}
\end{align}
%\Red{$C_m$ from metrics.}
Each of the $C_m$'s represents a possibility to extend $\fg^+$ with a single Virasoro generator, and obtain a Lie algebra with a non-singular shifted metric $\eta^{(m)}$, and can be written
\begin{align}
C_m=-\eta^{(m)\hat\alpha\hat\beta}t_{\hat\alpha}\otimes t_{\hat\beta}
\end{align}
using the respective metric.
Note that they act on tensor products of highest weight fundamentals; they carry index structure $(C_m)_{MN}{}^{PQ}$.
They manifestly have the property $\varsigma C_m\varsigma=C_m$, where $\varsigma$ is the permutation operator,
\ie, $(C_m)_{MN}{}^{PQ}=(C_m)_{(MN)}{}^{(PQ)}+(C_m)_{[MN]}{}^{[PQ]}$.
The commutators between $C_m$'s and the action of $t_\alpha$ induced by tensor product are
\begin{align}
[C_m,2(1\vee t_\alpha)]=\delta_\alpha^0mC_{m}+\delta_\alpha^1(m-1)C_{m+1}\;.
\label{CmVir10}
\end{align}
reflecting the invariance under the $\fg^+$ subalgebra. 

The tensor product of two fundamentals contains an infinite number of irreducible modules organised in a (finite) number of modules of the coset Virasoro algebra. The details of these modules of course depend on the central charge.
A universal property (independent of $\fg$) is that the leading symmetric module, which is the highest weight state in the corresponding coset Virasoro module, is annihilated by $C_m$, $m\geq-1$, and that the leading antisymmetric module is
annihilated by $C_0-2$ and by $C_m$, $m\geq1$.

%A way of expressing eq. \eqref{CmEq} is
%\begin{align}
%C_m=-\eta^
%\end{align}

%Coset Virasoro modules. Some universal features of these modules... Leading modules in t.p., section constraint.

The commutators of the $C_m$'s with mode shifted generators are
\begin{align}
[C_m,1\vee t_\alpha^{(n)}]={1\over2}\delta_\alpha^0(m-n)C_{m+n}+{1\over2}\delta_\alpha^1(m-n-1)C_{m+n+1}\;.
\label{CmVir1}
\end{align}
In many calculations, also the antisymmetrised product $1\wedge t_\alpha$ will be needed. This alone does not produce a result with a nice form. However, differences of such commutators obey
\begin{align}
[C_m,1\wedge t_\alpha^{(n)}]-[C_{m-q},1\wedge t_\alpha^{(n+q)}]=2q(1\wedge t_\alpha^{(m+n)})\;,
\label{ASCtCommutator}
\end{align}
which is straightforwardly derived from the explicit form \eqref{CmEq} of the $C_m$'s. 
This equation is a key to many of our calculations.

%Maybe something about how well defined operators are, \eg\ $G^{-1}C_m$...

%The 
As stated above, the coset Virasoro generators %can be seen as invariant operators, but 
are invariant only under the centrally extended loop algebra $\fg^+$. Under $L_0$ and $L_1$ they transform, see eq. \eqref{CmVir1}. The $L_0$ transformation is just by weight, which can still be considered covariant, but the $L_1$ transformation is non-trivial.
So, the operators $C_m$ are invariant if they are used with flat indices. 
With a covariantly constant vielbein $E$ at hand, covariantly constant operators $\tilde C_m$ can be constructed as
\begin{align}
\tilde C_m=(E\otimes E)C_m(E^{-1}\otimes E^{-1})\;.\label{tildeCmEq}
\end{align}
Note that only the Virasoro generators in $\fp$ are effective in this operation.
For example,%\Red{Explicit expressions.}
\begin{align}
\tilde C_0&=C_0-\phi C_1\;,\nn\\
\tilde C_1&=\varrho C_1\;,\label{tildeCExamples}\\
\tilde C_2&=\varrho^{2}\sum_{k=0}^\infty\phi^kC_{2+k}\;.\nn
\end{align}

\section{Affine extended geometry}

\subsection{The teleparallel complex\label{ComplexSec}}

The teleparallel formulation of extended geometry has been developed in refs. \cite{Cederwall:2021xqi,Cederwall:2023xbj}.
We find certain advantages to this version over the traditional coset dynamics.
Neither is fully geometric, in the sense that both diffeomorphisms and local $\fk(\fg^+)$ transformations are manifest in a tensorial way. The teleparallel formulation has the advantage that fields of all ghost numbers are identified from a tensor hierarchy algebra, and fit into a complex which is given the structure of Batalin--Vilkovisky theory \cite{Batalin:1981jr}, \ie, dual to an $L_\infty$ algebra
\cite{Lada:1992wc,Zwiebach:1992ie,Hohm:2017pnh,Roytenberg:1998vn}.
From a more practical point of view, we expect the formalism to be the ideal starting point for gauged supergravity and consistent truncations \cite{Bossard:2022wvi,Bossard:2023jid}.

One advantage with the problem of finding the correct dynamics in the teleparallel framework is that the problem has a clear-cut 
algebraic formulation, which in addition is present in its full form already in the linearised theory. This applies also to pure gravity. It is a matter of finding the $1$-bracket of a complex ${\mathscr C}$. We will sketch the structure here, and refer to ref. \cite{Cederwall:2023xbj} for details.
\begin{equation}\label{Cdiagram}
%\[
    \begin{tikzcd}[row sep = 16 pt, column sep = 16 pt]
    \hbox{ghost\#}=&2&1&0&-1&-2&-3\\
        \cdots\ar[r,"d"]&V'\ar[r,"d"]&V\ar[r,"d"]&\hat\fg\ar[r,"d"]&\Theta\ar[r,"d"]&\fk\\
        &&\bar\fk\ar[r,"d",swap]\ar[ur,"\rho" near start]&\bar\Theta\ar[r,"d",swap]\ar[ur,"\sigma" near start] 
        &\bar{\hat\fg}\ar[r,"d",swap]\ar[ur,"\rho^\star" near start]&\bar V\ar[r,"d",swap]&\bar V'\ar[r,"d",swap]&\cdots
%        \cdot & 10 \ar[drrr,controls={+(1,-1) and +(-1,1)}] 
 %       \cdot & \cdot & \cdot &\overline{16} & 10\ar[dr,shorten=-1mm]\\
\end{tikzcd}
%\]
\end{equation}
The fields in eq. \eqref{Cdiagram} are arranged so that the horizontal position is ghost number as indicated, and 
so that ghost number plus dimension (powers of inverse length) is $0$ in the upper line and $-1$ in the lower line.
Let us first describe the content of the upper line of the complex.

Linearised physical fields are found at ghost number $0$ as elements of the Lie algebra $\hat\fg=\fp\lplus\fg^+$.
Generalised diffeomorphisms are in $V$ (vectors),  $V'$ houses reducibilities of these, etc.
Antifields in the torsion modules are found in the vector space denoted $\Theta$, while $\fk$ contains some Bianchi identity modules projected on the compact subalgebra as ghost antifields.
These $(\fp\lplus\fg^+)$-modules are, at least for finite-dimensional structure algebras, obtained from the level decomposition of the tensor hierarchy algebra $S(\fg^{++})$ 
as described in Section \ref{FieldsFromTHASec}
\cite{Cederwall:2019bai}.

 The form of the $1$-bracket in the upper line (except the rightmost arrow) can also be obtained as a derived bracket. We will forego most of this translation (but see \eg\ eq. \eqref{DerivedGenDiffEq}).
In fact, the structure shown in \eqref{Cdiagram} is simplified, in that ancillary elements are left out. Ancillary fields arise in order to cancel local cohomology of the $1$-derivative $1$-bracket just described. Although essential for a correct description of the theory (for example, ancillary transformations arise in commutators of generalised diffeomorphisms in a generic case), most of the structure is not explicitly needed for the derivation of the dynamics of the physical fields.

A peculiar situation arises in the present case of (extended) affine structure algebra. As we will see in Section \ref{TorsionSection}, the set of torsion modules is larger than predicted by the tensor hierarchy algebra. Accordingly, also the set of Bianchi identities
(Section \ref{BISection}) is larger. We do not know if this is due to the singular properties of affine algebras, or if the ``enhancement'' will persist for further extended algebras. (Neither are we aware of any version of the tensor hierarchy algebra containing these modules, but suspect that one may exist.)
This affects the modules $\Theta$, $\bar\Theta$ in \eqref{Cdiagram} and the set of ancillary fields (suppressed in 
\eqref{Cdiagram}), not the existence of the complex, which probably necessitates the additional modules.

The upper line is mirrored in the lower line, where conjugate modules (with respect to integration, so in fact densities) are acted on by the natural dual of the $1$-bracket of the upper line.
The remaining parts of the $1$-bracket (disregarding ancillary fields) are $\rho$, the embedding of $\fk$ in $\hat\fg$
(defined by a background vielbein), its dual $\rho^\star$  and the ``dualisation'' $\sigma$. The latter is the true unknown, barring the question of which combination of Bianchi identities should be used for the last horizontal arrow in the upper line.
In order for this to be a complex, the $1$-bracket $q=d+\rho+\sigma+\rho^\star$ has to satisfy $q^2=0$, which leads to the condition
\begin{align}
d\rho+\sigma d=0\;.\label{dComplexEq}
\end{align} 
This should be seen as an algebraic condition on $\sigma$, which must be constructed as a sum of multiples of the identity on the $\fk$-modules in $\Theta$. 
With the natural pairing $\langle\cdot,\cdot\rangle$ on the complex, a linearised BV action for $\Psi\in{\mathscr C}$ is written as
\begin{align}
S_0={1\over2}\langle\Psi,q\Psi\rangle\;.
\end{align}
The form of $\sigma$ is unchanged in the non-linear theory.

Elimination of fields occurring algebraically (in the modules $\Theta$, $\fk$ and their duals) amounts to homotopy transfer to the cohomology of the diagonal arrows, resulting in the system
\begin{equation}
    \begin{tikzcd}[row sep = 16 pt, column sep = 16 pt]
       \cdots\ar[r]&V'\ar[r]&V\ar[r]&\hat\fg\ominus\fk\ar[dr,out=0,in=180,looseness=4]\\
 %       \cdots\ar[r]&V'\ar[r]&V\ar[r]&\fg\ominus\fh\ar[dr,to path={ -- ([xshift=2ex]\tikztostart.east) /--/
  %      								([xshift=-2ex]\tikztotarget.west) -- (\tikztotarget)}] \\
         &&&&\overline{\hat\fg\ominus\fk}\ar[r]&\bar V\ar[r]&\bar V'\ar[r]&\cdots
%        \cdot & 10 \ar[drrr,controls={+(1,-1) and +(-1,1)}] 
 %       \cdot & \cdot & \cdot &\overline{16} & 10\ar[dr,shorten=-1mm]\\
\end{tikzcd}
\end{equation}	
with a $2$-derivative equation of motion. This is the linearisation of the standard ``coset formulation'' of extended geometry.
Elimination of only $\Theta$, $\bar\Theta$, \ie, transfer to the cohomology of $\sigma$, leads to the system
\begin{equation}\label{RedComplex}
%\[
    \begin{tikzcd}[row sep = 16 pt, column sep = 16 pt]
%    \hbox{ghost\#}=&2&1&0&-1&-2&-3\\
        \cdots\ar[r,"d"]&V'\ar[r,"d"]&V\ar[r,"d"]&\hat\fg\ar[dr,out=0,in=180,looseness=3,"d\sigma^{-1}d", near end,swap]&&\fk \\
        &&\bar\fk\ar[ur,"\varrho" near start]&
        &\bar{\hat\fg}\ar[r,"d",swap]\ar[ur,"\varrho^\ast",near end,swap]&\bar V\ar[r,"d",swap]&\bar V'\ar[r,"d",swap]&\cdots
%        \cdot & 10 \ar[drrr,controls={+(1,-1) and +(-1,1)}] 
 %       \cdot & \cdot & \cdot &\overline{16} & 10\ar[dr,shorten=-1mm]\\
\end{tikzcd}
\end{equation}
This is the linearisation of the teleparallel formulation of extended geometry, with a kinetic term that is quadratic 
in torsion, contracted by a matrix $\sigma^{-1}$, \ie,
\begin{align}
S_0={1\over2}\langle \Theta(E),\sigma^{-1}\Theta(E)\rangle\;,
\label{ThetaSquareAction}
\end{align}
$\Theta(E)=dE$ being the linearised torsion.
Invariance under $\fk$ now follows from eq. \eqref{dComplexEq} as
$d\sigma^{-1}d\varrho=-d\sigma^{-1}\sigma d=-d^2=0$.
The step to the non-linear theory amounts to covariantisation. The only other issue in going from the linear to the non-linear model is the restricted form of the Bianchi identities, see Section \ref{BISection}.

In Section \ref{ActionSection}, we will use this form of the dynamics to derive the action
for teleparallel affine extended geometry. The method we will use is the covariant (non-linear) version of 
the complex \eqref{RedComplex}. Roughly speaking, all arrows in this complex, except the ``curved'' two-derivative one, follow from the tensor hierarchy algebra, as we will review
in Section \ref{FieldsFromTHASec}. The covariant version of this is the construction of torsion in Section
\ref{TorsionSection}. 
The arrows in \eqref{Cdiagram} which then are unknown are $\sigma$ and the arrow $\Theta\rightarrow\fk$ (and its dual). The former encodes the dynamics, in the sense of eq. \eqref{ThetaSquareAction}. The latter, although not present in 
\eqref{RedComplex}, encodes which combination of Bianchi identities into $\fk$ are used in showing the invariance.
Concretely, the dual arrow $\bar\fk\rightarrow\bar\Theta$ can be solved as $d=-\sigma^{-1}d\varrho$, and 
the arrow $\Theta\rightarrow\fk$ is constructed as $d=-\rho^\star d\sigma^{-1}$.

\subsection{Fields from the tensor hierarchy algebra\label{FieldsFromTHASec}}

The fields appearing at different ghost numbers in the upper line of the complex of Section \ref{ComplexSec}
(also the ancillary ones)
are identified using a tensor hierarchy algebra. This becomes increasingly important when the structure algebra one starts from is infinite-dimensional. It may then not be obvious what local symmetries, fields, fields strengths and Bianchi identities should appear (including ancillary ones). The affine case treated in the present paper is the first step to infinite-dimensional structure algebras, and the tensor hierarchy algebra informs us that the algebra we should consider is larger, containing also $L_1$. When we continue to \eg\ over-extended Kac--Moody algebras, simple guessing becomes virtually impossible. Then, also \eg\ the generalised diffeomorphisms get enriched with modules beyond the lowest weight fundamental.

The principle is the following \cite{Cederwall:2019bai}: We first extend the affine Kac--Moody algebra $\fg^+$
to an over-extended Kac--Moody algebra $\fg^{++}$ by adding a node 
to the Dynkin diagram (connected to the affine node)
%connected to the affine node in the Dynkin diagram,
and then to the tensor hierarchy algebra $S(\fg^{++})$ by adding another node (connected to the previous one),
which is grey ($\otimes$), meaning that the corresponding diagonal entry in the Cartan matrix is zero,
and that the associated generators are fermionic.
The tensor hierarchy algebra is thus a Lie superalgebra, defined by generators associated to the nodes and relations involving a Cartan matrix that can be read off from the Dynkin diagram. The corresponding extensions of finite-dimensional Lie algebras are related to the Borcherds--Kac--Moody algebras used in \cite{Henry-Labordere:2002xau, Henneaux:2010ys}. We will not review the construction in detail, but refer to ref. \cite{Cederwall:2021ymp}.

One considers the double grading of $S(\fg^{++})$ with respect to the two added nodes,
which we here label $-1$ and $-2$, so that the corresponding additional generators
in the extension from $\fg^+$ to $S(\fg^{++})$ include $e_{-1}$ and $\epsilon_{-2}$, where 
$e_{-1}$ is bosonic and $\epsilon_{-2}$
fermionic.
In our conventions for the double grading, the degree $(p,q)$ is such that $e_{-1}$
%, the raising operator associated with the first added (bosonic) node in a Chevalley-like basis, 
is found at degree $(1,1)$ and $\epsilon_{-2}$ %, the fermionic generator of the second step, is 
at 
degree $(0,-1)$.
In Appendix \ref{THAApp}, relevant elements in this grading and their brackets are tabulated.

%%Some notation:
%The module at $(p,q)$ is dual to the one at $(-p,1-q)$.
%%The loop generators $T_{Am}$ and $\KK$ act as in affine modules, \ie,
%%... 
%The eigenvalue of $\KK$ at $(p,q)$ is $k=-p$.
%%The generator $\dd$ acts as $L_0$, but with mode number shifts
%All elements come in pairs, related by what we call lowering and raising 
%\cite{Cederwall:2018aab,Cederwall:2019qnw}.
%Lowering is defined as $A^\fl=-[A,\epsilon_{-2}]$, and raising so that $\fl\sh+\sh\fl=1$.
%%(Mention $\KK^\sh$...)
%

All fields, non-ancillary as well as ancillary, are found as modules of the Lie algebra at degree $(0,0)$ (which is $\fp\inplus\fg^+$) at specific degrees. %Some notation:
The module at $(p,q)$ is dual to the one at $(-p,1-q)$.
%The loop generators $T_{Am}$ and $\KK$ act as in affine modules, \ie,
%... 
The eigenvalue of $\KK$ at $(p,q)$ is $k=-p$.
%The generator $\dd$ acts as $L_0$, but with mode number shifts
All elements come in pairs, related by what we call lowering and raising 
\cite{Cederwall:2018aab,Cederwall:2019qnw}.
Lowering is defined as $A^\fl=-[A,\epsilon_{-2}]$, and raising so that $\fl\sh+\sh\fl=1$.
%(Mention $\KK^\sh$...)
Ghost number is $p+q$. Non-ancillary fields are at $q=0$, and span the subalgebra $W(\fg^+)$.
Ancillary fields are found at $q=1$. They are present at degree $(p,1)$ when the module at degree $(p+1,1)$ is larger than the one at $(p+1,0)$, \ie, when there is some module $R_{(p+1,1)}$ annihilated by $\fl$. It can then be written as $R^\fl_{(p+1,2)}$.
The ancillary field at $(p+1,1)$ is formed as $[(B_{p+1,2})_M,F^{\fl M}]$, $B$ being an element in
a module $R_{(p+1,2)}$ with an extra section-constrained index, and $F^{\fl M}$ is a basis element at $(-1,-1)$. A part of the 1-bracket of the extended geometry is $\fl$. More information about the structure of $S(\fg^{++})$ at $-2\leq p\leq 2$
is given in Appendix~\ref{THAApp}, including Table~\ref{HyperbolicSTableBasis}.

In affine extended geometry, we extract the field content from Table \ref{HyperbolicSTableBasis}.
The elements at degree $(1,0)$ are ghosts of generalised diffeomorphisms, a vector in $R(-\lambda)$. 
There are also ghosts for ancillary transformations at degree $(0,1)$. These can be formed as
$[B_M{}^NL^\sh_N,F^{\fl M}]$,
where the index $M$ on $B_M{}^N$ is section-constrained,
which means that $B_M{}^N$ together with a derivative satisfies the section constraint given below as a condition
on pairs of derivatives.
These ghosts are of course important as local symmetries, but not explicitly used for the construction in the present paper.
(The invariance under generalised diffeomorphisms implies invariance under ancillary transformations, arising in the commutator of generalised diffeomorphisms.)
At degree $(0,0)$ we find the (linearised) fields in the algebra $\fp\lplus\fg^+$, and at degree $(-1,1)$ the ancillary section-constrained field $\gamma^-_M\Phi^{\sh M}=[\gamma^-_M\pi^\sh,F^{\fl M}]$.
Antifields in the torsion modules $\theta_M$ and $\Theta^-_M$ with weight $1$ and $2$, respectively, are found at degree $(-1,0)$ and Bianchi identities at $(-2,0)$.

It is, in retrospect, interesting to observe that the torsion we will use to construct the dynamics contains more than what is found in the tensor hierarchy algebra, see Section \ref{TorsionSection}.

\subsection{Coordinates, derivatives and section constraint\label{SCSection}}

Extended coordinates belong to a lowest weight module $R(-\lambda)$ of the structure algebra, and derivatives to the conjugate, highest weight, module $R(\lambda)$. 
Derivatives are subject to a section constraint, which for extended geometry in general reads
\begin{align}
Y\ket\*\otimes\ket\*=0\;, \label{SCEq}
\end{align}
where \cite{Bossard:2017aae,Cederwall:2017fjm}
\begin{align}
Y=-\eta^{\alpha\beta}t_\alpha\otimes t_\beta+(\lambda,\lambda)-1+\varsigma\;
\label{YEq}
\end{align}
in a normalisation of roots and weights where long roots $\alpha$ satisfy $(\alpha,\alpha)=2$.
The derivatives in eq. \eqref{SCEq} can act on anything. Solutions of the section constraint, ``sections'', are linear subspaces $S$ of the minimal orbit in $R(\lambda)$ under the structure group.
It states that the product of any two vectors in the section $S$ lies only in the highest symmetric product $R(2\lambda)$ and in the highest antisymmetric product $R(2\lambda-\alpha)$, where $\alpha$ is the root dual to $\lambda$, in our case the affine root.

The affine algebra $\fg^+$ does not have a non-degenerate metric. The correct choice (yielding precisely the leading modules in the tensor product) is to extend by $L_0$. The extension has an invariant metric, and insertion into eq. \eqref{YEq} yields
$Y=C_0-1+\varsigma$.
The leading modules, with eigenvalues $1-\varsigma$ of $C_0$, are also of course annihilated by all positive coset Virasoro generators. The symmetric one is in addition annihilated by $C_{-1}$. 
We thus have
\begin{align}
&C_m\ket A\vee\ket B=0\;,\quad m\geq-1\;,\nn \\
&(C_m-2\delta_{m,0})\ket A\wedge\ket B=0\;,\quad m\geq0\;.  \label{SectionConstraint} 
\end{align}
for $\ket A,\ket B\in S$. 

We notice that if the section constraint holds, it also holds expressed in flat indices, and vice versa. Namely, let 
$\ket A,\ket B\in S$ and let
$E^{-1}\ket A$ and $E^{-1}\ket B$ be the corresponding vectors transforming under $\fk(\fg^+)$.
Then,
\begin{align}
C_m(E^{-1}\ket A\otimes E^{-1}\ket B)=(E^{-1}\otimes E^{-1})\tilde C_m\ket A\otimes\ket B\;.
\end{align}
Each $\tilde C_m$ is a linear combination of $C_n$, $n\geq m$ according to eq. \eqref{tildeCmEq}.
This proves the statement,
which can be understood as the invariance of the section constraint under the full structure group $\gG^+ \rtimes \gP$.

\subsection{Generalised diffeomorphisms\label{GenDiffSection}}

A generalised diffeomorphism takes the general form in extended geometry:
\begin{align}
\LL_\xi V_M=\xi^N\*_NV_M+\eta^{\alpha\beta}t_{\alpha M}{}^Qt_{\beta N}{}^P\*_P\xi^N V_Q
+w\*_N\xi^NV_M\;.
\end{align}
The first term is a transport term, the second and third ones transformations in the structure algebra and scaling.
We have chosen to display the action on a covector density $V$; other modules of the structure algebra follow.
The weight $w$ is arbitrary, the canonical value for a covector is $1-(\lambda,\lambda)$.
As a ``derived bracket'' from the underlying tensor hierarchy algebra (see Appendix \ref{THAApp}), 
this generalised Lie derivative is constructed as
\begin{align}
\LL_\xi A=[[\xi,F^{\fl M}],\partial_M A^\sh]-[[\*_M\xi^\sh,F^{\fl M}],A]\;,\label{DerivedGenDiffEq}
\end{align}
with $\xi=\xi^NE_N$,
when $A$ is some object obeying $A^\flat=0$.

In affine extended geometry, $(\lambda,\lambda)=0$. The correct choice for the inverse metric is again the one obtained from the extension of $\fg^+$ with $L_0$, leading to
\begin{align}
\LL_\xi V_M=\xi^N\*_NV_M-(C_0)_{MN}{}^{QP}\*_P\xi^N V_Q
+w\*_N\xi^NV_M\;.\label{WeightDef}
\end{align}
Note that, due to the metric component $(L_0,\KK)=-1$, weight
is shift of the action of $L_0$ (for a covector with a minus sign).
%If we define weight by eq. \eqref{WeightDef}, 

%\section{Teleparallel dynamics}

\subsection{Connection and torsion\label{TorsionSection}} 

The generalised vielbein $E_M{}^A$ is covariantly constant,
\begin{align}
D_ME_N{}^A=\*_ME_N{}^A+\Gamma_{MN}{}^PE_P{}^A=0\;,
\end{align}
 if the connection is chosen as the right-invariant Maurer--Cartan form
\begin{align}
\Gamma_{MN}{}^P=-(\*_MEE^{-1})_N{}^P\;.
\end{align}
This is the (generalised) Weitzenb\"ock connection. Note that the covariant constancy does not involve a spin connection.
The connection has a derivative for its first index, which means it is in section; it obeys the section constraint together with any other section-constrained object.
It fulfils the Maurer--Cartan equation
\begin{align}
\*_{M}\Gamma_{N}-\*_{N}\Gamma_{M}+[\Gamma_{M},\Gamma_{N}]=0\;.\label{MCeq}
\end{align}
The connection is thus flat, expressing teleparallelism.

%[{\bf G:B: This paragraph is not clear, $\gamma_M = - \Gamma_M{}^0$ is not defined and the definition of $\dd=L_0$ makes it all %strange.} 
In a generic case where the structure algebra is $\fg\oplus\RR$ for some (finite-dimensional) Lie algebra $\fg$, the parts of the connection corresponding to $\fg$ and $\RR$ are extracted as
$\Gamma_M=t_\alpha\Gamma_M{}^\alpha+w\gamma_M$,
where $w$ is the weight of $E$ and $\gamma_M$ the scaling connection.
In the present case, however, with affine structure algebra, we have seen that weight is encoded in the shifted action of $\dd = \ell_0-1$
(see \eg\ eq. \eqref{dWeightTransf}). 
%The last term in eq. \eqref{GammaDecompEq} is thus incorporated in the first term.
In covariant derivatives, $\Gamma_M{}^0$ will be accompanied by the representation matrix of $\dd$ of the object it acts on, including weight.
In order for the definition of the connection components to be independent of the weight assigned to 
$E=e^{-\phi\ell_1}\varrho^{-\ell_0+1}e$, we let
\begin{align}
\Gamma_M=\Gamma_M{}^\alpha t_\alpha-\Gamma_M{}^0\;.
\end{align}
Then, %\Red{fix first term!}
\begin{align}
\Gamma_M{}^\alpha t_\alpha =-e^{-\phi\ad{\ell_1}}\varrho^{-\ad{\ell_0}}(\partial_M e  e^{-1}) + {\partial_M\varrho\over\varrho}\ell_0 + \bigl( \partial_M\phi-\phi{\partial_M\varrho\over\varrho}\bigr) \ell_1  \;.
\end{align}
%Above, it is assumed that the vielbein carries a canonical weight 
%If so desired, the vielbein may be assigned a weight $w$, together with an extra term $

%In the bra-ket notation, choosing the canonical weight $1-(\lambda,\lambda)=1$ for the vielbein, the above equations become
%\begin{align}
%...
%\end{align}

Torsion is by definition the tensorial part of the connection.
From the tensor hierarchy algebra, we expect torsion to appear as a covector (as usual) together with a minus-shifted covector. The latter is what encodes the ``big'' torsion module appearing for finite-dimensional structure algebras, and the former the ``small'' one in $R(\lambda)$ (present in $W(\fg^+)$ but not in $S(\fg^+)$).
The inhomogeneous part of the transformation of the connection under generalised diffeomorphisms,
$\Delta_\xi\equiv\delta_\xi-\LL_\xi$, is
\begin{align}
\Delta_\xi\Gamma_M{}^\alpha&=-\eta^{(0)\alpha\beta}t_{\beta N}{}^P\*_M\*_P\xi^N\;,\nn\\
\hbox{\ie, } \Delta_\xi\ket{\Gamma^\alpha}&=-\eta^{(0)\alpha\beta}(1\otimes\bra\xi)(1\otimes t_\beta)\ket{\*_\xi}\otimes\ket{\*_\xi}\;.
\label{GammaInhomEq}
\end{align}
Subscripts on $\*$ indicate the object they act on.
If we form the fundamental 
\begin{align}
\ket\theta=t_\alpha\ket{\Gamma^\alpha}\;,\label{ThetaDef}
\end{align}
this leads to 
\begin{align}
\Delta_\xi\ket\theta=(1\otimes\bra\xi)C_0\ket{\*_\xi}\otimes\ket{\*_\xi}=0\;,
\end{align}
vanishing thanks to the symmetric section constraint.

The minus-shifted torsion should then tentatively appear as $t^-_\alpha\ket{\Gamma^\alpha}$ (this is also the result of the na\"\i ve $1$-bracket derived from the tensor hierarchy algebra). However, $\KK^-=0$, and we only obtain
\begin{align}
\Delta_\xi(t^-_\alpha\ket{\Gamma^\alpha})=
(1\otimes\bra\xi)\Bigl(\ell_{-1}\otimes 1-\sum_{n\in\ZZ}\eta^{AB}t_{A,n}\otimes t_{B,-1-n}\Bigr)
\ket{\*_\xi}\otimes\ket{\*_\xi}\;.
\end{align}
In order to get the missing term to build $C_{-1}$ and use the section constraint, we need also to introduce an ancillary {\it field}
$\ket{\gamma^-}$ with inhomogeneous transformation rule 
$\Delta_\xi\gamma^-_M=(\ell_{-1})_N{}^P\*_M\*_P\xi^N$, and let
\begin{align}
\ket{\Theta^-}=\ket{\gamma^-}+t^-_\alpha\ket{\Gamma^\alpha}\;.\label{ThetaMinusDef}
\end{align}
$\ket{\gamma^-}$ is section-constrained.
All this is precisely what is read off from the tensor hierarchy algebra, Section \ref{FieldsFromTHASec}, and agrees with the nilpotent derived $1$-bracket obtained from it (covariance of torsion is the non-linear version of the nilpotency of the 1-bracket starting at diffeomorphism ghosts).
Even if we do not find such torsion in the tensor hierarchy algebra, it is clear from the procedure that it can be repeated with any shift $n\geq-1$, so there is torsion
\begin{align}
\ket{\Theta^{(n)}}=\ket{\gamma^{(n)}}+t^{(n)}_\alpha\ket{\Gamma^\alpha}\;,\quad n\geq-1\;,
\end{align}
where $\ket{\gamma^{(0)}}=0$ and $\Delta_\xi\gamma^{(n)}_M=(\ell_n)_N{}^P\*_M\*_P\xi^N$, $n\geq -1$. The inhomogeneous part of variation under generalised diffeomorphism of $\ket{\Theta^{(n)}}$ then gives 
\begin{equation} \Delta_\xi \ket{\Theta^{(n)}} =(1\otimes\bra\xi)C_n \ket{\*_\xi}\otimes\ket{\*_\xi}=0 \; , \end{equation}
which vanishes for $n\geq-1$ according to the section constraint \eqref{SectionConstraint}. 

The non-covariance of the shifts necessitates checking the transformations of the shifted torsions under the Virasoro generator $L_1$.  Using, for any element $A$ in the Virasoro-extended affine algebra, that $s[L_1,A]=[L_1,sA]+s^2A$, iterated to
$s^n[L_1,A]=[L_1,s^nA]+ns^{n+1}A$, this implies the off-diagonal action of $L_1$ on shifted torsions:
\begin{align}
L_1\cdot\ket{\Theta^{(n)}}=\ell_1\ket{\Theta^{(n)}}+n\ket{\Theta^{(n+1)}}\;.
%{\mathbb L}_1=\left(\begin{matrix}
%\ell_1&-1\\
%&\ell_1\\
%&&\ell_1&1\\
%&&&\ell_1&2\\
%&&&&\ell_1&3\\
%&&&&&\ddots
%\end{matrix}\right)
\end{align}
%(zero entries suppressed).
This of course also applies to terms in covariant derivatives containing $\Gamma^1$.
Note that it is in agreement with the brackets in the tensor hierarchy algebra, see \eg\ eq.
\eqref{JordanFPhi}.
The off-diagonal elements of $L_1$ in this module can be eliminated by forming new field-dependent 
combinations\footnote{And also $\ket{\tilde\Theta^{(n)}}=\varrho^{n}\sum_{i=0}^\infty (^{-n}_{\; \; i}) (-\phi)^i\ket{\Theta^{(i+n)}}$ for $n\geq2$, which will not be used.} (with good properties also under $L_0$):
\begin{align}
\ket{\tilde\Theta^-}&=\varrho^{-1}(\ket{\Theta^-}-\phi\ket\theta);,\label{TildeThetaEq}\nn\\ 
\ket{\tilde\Theta^+}&=\varrho\sum_{i=0}^\infty\phi^i\ket{\Theta^{(i+1)}}
\end{align}
%{\Red{I don't think this expression is useful, but I corrected it. I would remove it}}
The covariant derivatives on $\ket{\tilde\Theta^\pm}$ fulfil 
\begin{align}
\ket D\otimes\ket{\tilde\Theta^-}&=\varrho^{-1}(\ket D\otimes\ket{\Theta^-}-\phi \ket D\otimes\ket\theta)\;,\label{xxxeq}\nn\\
\ket D\otimes\ket{\tilde\Theta^+}&=\varrho\sum_{i=0}^\infty\phi^i\ket D\otimes\ket{\Theta^{(i+1)}}\;.
\end{align}
%so the $\phi$ factors in the recombinations can be viewed as covariantly constant.

%where $s^{-1}$, negative mode-shift, is the inverse of $s$ on its cokernel (the kernel being spanned by $\KK$).
%(fix $\varrho$ dependence). 
%\Red{Notation is inefficient. Write $\tilde T^\pm$ as operations $\tilde s_\pm T$.}

A convenient way of deriving the $\tilde\Theta$'s is the covariant procedure of ``flattening'' indices with $E^{-1}$, then shifting indices, and finally reverting to coordinate basis indices with $E$. 
We have $E^{-1}\ket\Theta=E^{-1}t_\alpha E\ket{E^{-1}\Gamma^\alpha}$.
Acting on generators, we define
\begin{align}
\tilde T^+\equiv\tilde s T= Es(E^{-1}TE) E^{-1}\;.\label{TildeSEq}
\end{align} 
Note that under involution
\begin{align}
\tau(\tilde T^\pm)=\tau(\tilde T)^\mp\;.\label{yyyeq}
\end{align}
We can again make use of the action of $\gP$ of eq. \eqref{PonZAction}. 
We find, on any function $f(z)$,
\begin{align}
\tilde f^-(z)&\equiv ( e^{-\phi L_1}\circ\varrho^{-L_0}\circ s^{-1}\circ\varrho^{L_0}\circ e^{\phi L_1})f(z)
=\varrho^{-1}\left({1\over z}-\phi\right)f(z)\;,\\
\tilde f^+(z)&\equiv ( e^{-\phi L_1}\circ\varrho^{-L_0}\circ s\circ\varrho^{L_0}\circ e^{\phi L_1})f(z)
={\varrho\over{1\over z}-\phi}f(z)=\varrho\sum_{k=0}^\infty \phi^kz^{k+1}f(z)\;.\nn
\end{align}
We recognise the linear combinations from eq. \eqref{TildeThetaEq}. 
Thus,
\begin{align}
\ket{\tilde\Theta^{(m)}}=\ket{\tilde \gamma^{(m)}}+E(E^{-1}t_\alpha E)^{(m)}\ket{E^{-1}\Gamma^\alpha}\;.
\label{TildeThetaMM}
\end{align}
The relations \eqref{xxxeq} and \eqref{yyyeq} are now manifest.

In conclusion, we have found the torsion $\Theta^-$ and $\theta$ with weights $2$ and $1$, respectively, predicted by the tensor hierarchy algebra, but also 
$\Theta^{(n)}$, $n>0$, with weight $1-n$. The field-dependent combinations that will be used in the construction of the action are
$\tilde\Theta^{(n)}$, $n=-1,0,1$, all with weight $1$. Of these, only the first two can be formed from the ones in the tensor hierarchy algebra.

\subsection{Weights and integration\label{WeightsSec}}

An object with weight $w$, transforming as eq. \eqref{WeightDef} under generalised diffeomorphisms (with the obvious extension to arbitrary $\fg^+$ modules) is acted on with the covariant derivative \begin{align}
D_MV_N=\*_MV_N+\Gamma_M{}^\alpha t_{\alpha N}{}^PV_P-w\Gamma_M{}^0V_N\;.
\end{align}
The last term arises since the inhomogeneous transformation \eqref{GammaInhomEq} gives 
$\Delta_\xi\Gamma_M{}^0=\*_M\*_N\xi^N$.
%With the parametrisation $E=e^{-\phi L_1}\varrho^{-L_0+1} e$, the Virasoro connections are
%\begin{align}
%\Gamma^0&={d\varrho\over\varrho}\;,\nn\\
%\Gamma^1&=d\phi-\phi{d\varrho\over\varrho}\;.\label{01Connections}
%\end{align} 

The scale parameter $\varrho$ is a covariantly constant scalar density of weight $1$ (thus subsuming the r\^ole of the determinant of the metric).
%$\phi$ is...
As is also seen from the list of weights in the tensor hierarchy algebra,  Table \ref{HyperbolicSTableBasis}, $(p,q)=(-1,0)$, the weight of the unshifted torsion $\ket\theta$ is $1$, the tensorial value. Shifted torsion $\ket{\Theta^{(n)}}$ carries weight
$1-n$. All $\ket{\tilde\Theta^{(n)}}$ are tensors, and have weight $1$.

If $V^M$ is a vector density with weight $0$, 
\begin{align}
D_MV^M=\*_MV^M-\Gamma_M{}^\alpha t_{\alpha N}{}^MV^N=(\*_M-\theta_M)V^M\;.
\end{align}
So a naked divergence is covariant, and $\*_MV^M=(D_M+\theta_M)V^M$ is a scalar density with weight 1.

This is relevant for showing invariance, modulo a total derivative, of an action under local rotations in $\fk(\fg^+)$.
Let the Lagrangian density be $\LL$ (of weight $1$), and its variation $D_M\Upsilon^\alpha Z_\alpha{}^M$, where $\Upsilon$ is the variation parameter.\footnote{The variation parameter $\tilde{\Upsilon} \in \fk(\fg^+)$ is defined such that $\delta E = E \tilde{\Upsilon} = \Upsilon E$ and $\delta \Gamma_M = - E (\partial_M \tilde{\Upsilon} )E^{-1} = - D_M \Upsilon$.} Then,
\begin{align}
\delta_\Upsilon\LL=D_M\Upsilon^\alpha Z_\alpha{}^M=\*_M(\Upsilon^\alpha Z_\alpha{}^M)
-\Upsilon^\alpha(D_M+\theta_M)Z_\alpha{}^M\;.\label{PartIntEq}
\end{align}
If we want the second term to vanish with the help of Bianchi identities, it follows that these must contain $\bar D\equiv D+\theta$.
In terms of the teleparallel complex, the same observation follows from the duality under integration of the $1$-brackets in a given background \cite{Cederwall:2023xbj}.

\subsection{Bianchi identities\label{BISection}} 

It is clear already from the complex in Section \ref{ComplexSec} that Bianchi identities are responsible for the 
invariance under local rotations in teleparallel theories. 

In the light of the considerations in Section \ref{WeightsSec}, we are interested specifically in Bianchi identities which can be written on the form
\begin{align}
(D+\theta)_PZ_{MN}{}^P=0\;,
\end{align}
with $Z$ being linear in torsion, and where the pair $MN$ may be converted into an adjoint index.
%This is because $\bar D\equiv D+\theta$ arises as the result of a partial integration. ß
All Bianchi identities do not have this form. From the example calculation in Appendix \ref{BianchiProofApp}, it is seen that the presence of $\theta$ actually simplifies many terms and makes the proof easier.
We have a conjecture that Bianchi identities of this type appear in the tensor hierarchy algebra as modules which are not in the image of $\flat$, and expect that this may be proven using the methods of refs. 
\cite{Cederwall:2022oyb,Cederwall:2023stz}.

A new feature, compared to finite-dimensional structure algebras, is that not only antisymmetric Bianchi identities, but also symmetric ones, enter the proof that the action is invariant modulo a total derivative under local rotations. 

The following is a list of Bianchi identities of the desired form, expressed in terms of shifted torsion, that will be used for showing invariance of the action.
\begin{align}
&C_1\ket{\bar D }\wedge\ket{\Theta^-}-(C_0+2)\ket{\bar D }\wedge\ket{\theta}=0\;,\nn\\
&2C_2\ket{\bar D }\wedge\ket{\Theta^{(n)}}-3C_1\ket{\bar D }\wedge\ket{\Theta^{(n+1)}}
+(C_0-2)\ket{\bar D }\wedge\ket{\Theta^{(n+2)}}=0\;,\quad n\geq-1\;,\nn\\
&C_1\ket{\bar D }\vee\ket{\Theta^{(n)}}-C_0\ket{\bar D }\vee\ket{\Theta^{(n+1)}}=0\;, \quad n\geq-1\;,\label{BIEq}\\
&C_m\ket{\bar D }\otimes\ket{\Theta^{(n)}}-2C_{m-1}\ket{\bar D }\otimes\ket{\Theta^{(n+1)}}
+C_{m-2}\ket{\bar D }\otimes\ket{\Theta^{(n+2)}}=0\;, \quad m\geq3\;,n\geq-1\;.\nn
\end{align}
%where $\ket{\bar D }=\ket{D+\theta}$.
Only the first identity uses the Maurer--Cartan equation for the connection, the other ones are more ``trivial''. All of course rely on the section constraint.
There is also a Bianchi identity
\begin{align}
(C_0-2)\ket{\bar D }\wedge\ket{\Theta^-}-C_{-1}\ket{\bar D }\wedge\ket{\theta}\;,\label{UnnecBI}
\end{align}
which is not used in the construction of the action.
We refer to Appendix \ref{BianchiProofApp} for the method of proof of these Bianchi identities, and one explicit example.

The Bianchi identities that we eventually will use involve $\tilde\Theta^\pm$, not $\Theta^\pm$.
We will now demonstrate that Bianchi identities of the same form as above hold, if all $C_m$ are replaced by $\tilde C_m$ and
$\Theta^{(m)}$ by $\tilde\Theta^{(m)}$, and vice versa. 
%This may sound peculiar, since in the relation between $C$'s and $\tilde C$'s there was a conversion between curved and flat indices

Consider a term $\tilde C_m\ket D\otimes\ket{\tilde\Theta^{(n)}}$.
We can now use eq. \eqref{TildeThetaMM} to write it as
\begin{align}
\tilde C_m\ket D\otimes\ket{\tilde\Theta^{(n)}}=(E\otimes E)C_m\ket{E^{-1}D}\otimes(\ket{E^{-1}\tilde \gamma^{(n)}}+(E^{-1}t_\alpha E)^{(n)}\ket{E^{-1}\Gamma^\alpha})\;.\label{CDThetaEq}
\end{align}
Furthermore, let $E^{-1}t_\alpha E=\xi_\alpha{}^\beta t_\beta$ and $\Gamma'^\alpha=\xi_\beta{}^\alpha\Gamma^\beta$. The first of these equations may look confusing, since it seems to equate a left hand side carrying flat indices with a right hand side carrying coordinate basis indices, but makes sense by just thinking of $E$ as a group element, then $\xi^{-1}$ is the corresponding group element in the adjoint representation, and the equation states invariance of representation matrices. Then,
\begin{align}
\tilde C_m\ket D\otimes\ket{\tilde\Theta^{(n)}}
=(E\otimes E)C_m\ket{E^{-1}D}\otimes(\ket{E^{-1}\tilde \gamma^{(n)}}+t_\alpha^{(n)}\ket{E^{-1}\Gamma'^\alpha})\;.
\end{align}
Since $\xi$ is a conjugation, $\Gamma'$ obeys the same Maurer--Cartan equation as $\Gamma$. The other ingredients in the derivation of the Bianchi identities \eqref{BIEq} are the commutators between $C_m$ and $t^{(n)}_\alpha$ and the section constraint. The shifted generators now stand next to $C_m$, just like in $C_m\ket D\otimes\ket{\Theta^{(n)}}$, and the section constraint holds also for flattened indices (Section \ref{SCSection}). Therefore, the same proof of Bianchi identities holds as in 
Appendix \ref{BianchiProofApp}. The argument also goes the other way.

The proof is abstract enough to warrant some illustration.
Consider, as an example, the antisymmetric covariant Bianchi identity
\begin{align}
2\tilde C_2\ket{\bar D}\wedge\ket{\tilde\Theta^-}-3\tilde C_1\ket{\bar D}\wedge\ket{\theta}
+(\tilde C_0-2)\ket{\bar D}\wedge\ket{\tilde\Theta^+}=0\;.
\end{align}
We will expand the left hand side in $\phi$ using the relations \eqref{TildeThetaEq} and \eqref{tildeCExamples}.
This gives
\begin{align}
&2\tilde C_2\ket{\bar D}\wedge\ket{\tilde\Theta^-}-3\tilde C_1\ket{\bar D}\wedge\ket{\theta}
+(\tilde C_0-2)\ket{\bar D}\wedge\ket{\tilde\Theta^+}\nn\\
&=\varrho\sum_{k=0}^\infty\phi^k\Bigl(\Bigr.
2C_{2+k}\ket{\bar D}\wedge\ket{\Theta^-}-2C_{1+k}\ket{\bar D}\wedge\ket\theta\\
&\qquad\qquad\qquad-C_1\ket{\bar D}\wedge\ket{\Theta^{(k)}}+C_0\ket{\bar D}\wedge\ket{\Theta^{(1+k)}}
\Bigl.\Bigr)\;.\nn
\end{align}
The 4-term sum inside the parenthesis for a given $k$ can, by repeatedly adding and subtracting the same terms, be written as
\begin{align}
&2\sum_{\ell=0}^{k-1}\left(
C_{2+k-\ell}\ket{\bar D}\wedge\ket{\Theta^{(-1+\ell)}}-2C_{1+k-\ell}\ket{\bar D}\wedge\ket{\Theta^{(\ell)}}
+C_{k-\ell}\ket{\bar D}\wedge\ket{\Theta^{(1+\ell)}}
\right)\nn\\
&+2C_{2}\ket{\bar D}\wedge\ket{\Theta^{(-1+k)}}-3C_{1}\ket{\bar D}\wedge\ket{\Theta^{(k)}}
+(C_{0}-2)\ket{\bar D}\wedge\ket{\Theta^{(1+k)}}\;,
\end{align}
which vanishes thanks to the ``untilded'' Bianchi identities.
	
The Bianchi identities that will be used in Section \ref{ActionSection} are those involving only $\tilde\Theta^-$, $\theta=\tilde\Theta^{(0)}$ and 
$\tilde\Theta^+$. They read:
\begin{align}
&\tilde C_1\ket{\bar D }\wedge\ket{\tilde \Theta^-}-(\tilde C_0+2)\ket{\bar D }\wedge\ket{\theta}=0\;,\nn\\
&2\tilde C_2\ket{\bar D }\wedge\ket{\tilde \Theta^-}-3\tilde C_1\ket{\bar D }\wedge\ket{\theta}
+(\tilde C_0-2)\ket{\bar D }\wedge\ket{\tilde \Theta^+}=0\;,\nn\\
&\tilde C_1\ket{\bar D }\vee\ket{\tilde \Theta^-}-\tilde C_0\ket{\bar D }\vee\ket{\theta}=0\;,\label{tildeBIEq}\\
&\tilde C_1\ket{\bar D }\vee\ket{\theta}-\tilde C_0\ket{\bar D }\vee\ket{\tilde \Theta^+}=0\;,\nn\\
&\tilde C_m\ket{\bar D }\otimes\ket{\tilde \Theta^-}-2\tilde C_{m-1}\ket{\bar D }\otimes\ket{\theta}
     +\tilde C_{m-2}\ket{\bar D }\otimes\ket{\tilde \Theta^+}=0\;, \quad m\geq3\;.\nn
\end{align}
Only the first and third of these identities contain torsion within the tensor hierarchy algebra, and are present in it at level
$(p,q)=(-2,0)$.

\subsection{Teleparallel affine geometry\label{ActionSection}}

%\subsubsection{General Ansatz}

We search the explicit form of the ``kinetic operator'' $\sigma^{-1}$ (see Section \ref{ComplexSec}), such that the Lagrangian
%\footnote{The overall factor $\varrho$ assumes the use of a tensorial vielbein. If $E$ carries weight 0, the factor will instead be $\varrho^{-1}$.}
\begin{align}
\LL={1\over2}\varrho(\sigma^{-1})^{MN}_{(m,n)}\Tilde\Theta^{(m)}_M\Tilde\Theta^{(n)}_N
={1\over2}\varrho\brabra{(\sigma^{-1})_{(m,n)}} \ket{\tilde\Theta^{(m)}}\otimes\ket{\tilde\Theta^{(n)}}
\label{FirstLEq}
\end{align}
is invariant under local $\gK(\gG^+)$ rotations. Summation over $m,n=-1,0,1$ is understood.
The overall factor $\varrho$ gives $\LL$ weight $1$ (all other objects are tensors), appropriate for integration.

%\Red{Some reasoning about why $C_2$ should be present etc...Would fit in Section \ref{PhysicalSection}}

The tools at hand for constructing the kinetic operator are the invariant tensors of Section \ref{InvariantTensorsSection} together with the metric $G$. We assume, without any restriction, that\footnote{All ket vectors carry indices that come from derivatives, \ie, from section-constrained covectors.  In a suitable basis, these are of finite depth, and contraction with $G^{-1}$ is well defined.}
$\brabra{(\sigma^{-1})_{(m,n)}}=\brabra{G^{-1}}\tilde X_{(m,n)}$,
 so that the Ansatz \eqref{FirstLEq} can be written as
\begin{align}
\LL={1\over2}\varrho\brabra{G^{-1}}\tilde X_{(m,n)}\ket{\tilde\Theta^{(m)}}\otimes\ket{\tilde\Theta^{(n)}}\;.
\label{XLagrangian}
\end{align}
Each $\tilde X_{(m,n)}$ is some element in $\End(\otimes^2F)$, acting on the tensor product of (shifted) fundamentals. In the end, these will be linear combinations of the identity operator and coset Virasoro operators $\tilde C_m$.
We will present (candidate) terms in $\tilde X$ as matrices; it should then be remembered that each entry is an operator.
%Here we reverted to the index-free notation of Section \ref{NotationSection}; the 
%expression $\brabra{G^{-1}}M\ket A\otimes\ket B$ is in index notation read $G^{MN}M_{MN}{}^{PQ}A_PB_Q$. 
Since all $(G^{-1}\tilde X)^{MN}$ are symmetric in $(MN)$, we must have $\tilde X_{(m,n)}=\varsigma\tilde X_{(n,m)}\varsigma$.
Since the identity and the coset Virasoro generators fulfil ${\mathscr O}=\varsigma{\mathscr O}\varsigma$, we arrive at
$\tilde X_{(m,n)}=\tilde X_{(n,m)}$.
Letting the corresponding operators in flat indices be 
$X_{(m,n)}=(E^{-1}\otimes E^{-1})\tilde X_{(m,n)}(E\otimes E)$, the Lagrangian is written as
\begin{align}
\LL={1\over2}\varrho\brabra{H^{-1}}X_{(m,n)}(E^{-1}\otimes E^{-1})\ket{\tilde\Theta^{(m)}}\otimes\ket{\tilde\Theta^{(n)}}\;.
\label{LwithHEq}
\end{align}

The transformations under local rotations with parameter $\Upsilon\in\fk(\fg^+)$ is, as already stated, 
$\delta_\Upsilon E=-E\Upsilon$, leading to
\begin{align}
\delta_\Upsilon\tilde\Theta^{(m)}=D\tilde s^{(m)}(E\Upsilon E^{-1})=D(Es^m\Upsilon E^{-1})
\end{align}	
where $\tilde s$ is defined by \eqref{TildeSEq}.
 After partial integration, using eq. \eqref{PartIntEq}, this leads to terms\footnote{Here, it is essential that $\tilde X$ is covariantly constant, which it is when built from coset Virasoro generators with {\it flat} indices and vielbeins, \ie, from $\tilde C_m$'s.}
 \begin{align}
(\delta_\Upsilon\LL)=\varrho\brabra{H^{-1}}X_{(m,n)}(\Upsilon^{(m)}\otimes1)(E^{-1}\otimes E^{-1})\ket{\bar D }\otimes\ket{\tilde\Theta^{(n)}}
\label{DeltaLambdaX}
\end{align}
module a total derivative,
where $\ket{\bar D  }=\ket{D+\theta}$ acts on the object to its right, $\ket{\tilde\Theta^{(n)}}$.

There are two essential pieces of information that may be used in order to show invariance:
the property $\Upsilon\in\fk(\fg^+)$ and the Bianchi identities.
%\footnote{We know on general grounds that the properties of the kinetic operator relies on purely algebraic relations, and does not use \eg\ the section constraint.}.
The former is expressed as $\tau_H(\Upsilon)=\Upsilon$ (this is the Chevalley involution). Furthermore, using the general statement
$\brabra{H^{-1}}(1\otimes T)=-\brabra{H^{-1}}(\tau_H(T)\otimes1)$, we can make use of eq. \eqref{ModeInvolutionOne} to obtain
\begin{align}
\brabra{H^{-1}}(\Upsilon\vee1)&=0\;,\nn\\
\brabra{H^{-1}}(\Upsilon^-\otimes1)&=\brabra{H^{-1}}(\Upsilon^+\wedge1)-\brabra{H^{-1}}(\Upsilon^+\vee1)\;.
\label{GLambdaIds}
\end{align}

Before these identities can be set to work, and also before Bianchi identities can be extracted, the coset Virasoro 
generators in $X_{(m,n)}$ must be moved past $(\Upsilon^{(m)}\otimes1)$ in eq. \eqref{DeltaLambdaX}.
We thus need to deal with commutators $[X_{(m,n)},\Upsilon^{(m)}\otimes1]$.
Since $\Upsilon$ lies in the loop algebra, the symmetric product commutes,
$[X_{(m,n)},\Upsilon^{(m)}\vee1]=0$. For the antisymmetric product, we need to rely on eq.
\eqref{ASCtCommutator}.

Considering eq. \eqref{DeltaLambdaX}, and only the terms linear in coset generators, 
we see that the rows of the matrix $X_{(m,n)}$ need to be such that eq. \eqref{ASCtCommutator} can be applied, \ie, the rows must be sums of neighbouring $(C_m,-C_{m-1})$, and that the columns need to provide (non-constant terms in) Bianchi identities. In addition, if this happens already before 
eq. \eqref{GLambdaIds} is applied, an expression is trivial (or reduces to an expression without coset generators).

This puts strong restrictions on the matrices that can be used, but we need a starting point.
One piece of information can be gained from reducing to adjoint extended geometry.
This is done by choosing a solution of the affine section constraint where 
$\ket\partial=\ket0\partial_0+t_{A,-1}\ket0\partial^A$, with the adjoint $\partial^A$ satisfying the section constraint of adjoint extended geometry, and restricting the vielbein to an adjoint group element.
We will not perform the complete reduction. The only input we need here is the quite straightforwardly derived observation that 
there must be a term 
 $\brabra{G^{-1}}\tilde C_2\ket{\tilde\Theta^-}\otimes\ket{\tilde\Theta^-}$.
 This can also be deduced by a comparison to the dynamics in the coset formulation
 \cite{Bossard:2018utw}.
 
In order for the rows to fulfil the above requirement (with a symmetric matrix), this necessitates
a term in $X$
\begin{align}
N=\left(\begin{matrix}C_2&-C_1&0\\-C_1&C_0&0\\0&0&0\end{matrix}\right)\;,
\end{align}
where rows and columns are labelled by $m,n=-1,0,1$.
However, while the second column corresponds to symmetric and antisymmetric Bianchi identities (there will be remainders, with the identity matrix), the first column does not. The matrix $N$ must be combined with something else in order to obtain Bianchi identities, after using eq. \eqref{GLambdaIds}. The key property is in fact that the symmetric and antisymmetric Bianchi identities (second and third equations in \eqref{BIEq}) contain different combinations of coset Virasoro generators, with the right combinations arising from the different signs for 
$(\Upsilon^+\wedge1)$ and $(\Upsilon^+\vee1)$ in eq. \eqref{GLambdaIds}.
Another building block is
\begin{align}
M=\left(\begin{matrix}C_4&-2C_3&C_2\\-2C_3&4C_2&-2C_1\\C_2&-2C_1&C_0\end{matrix}\right)\;.
\end{align}
For this matrix, the first two columns give symmetric and anti-symmetric Bianchi identities, while the last column does not.

Note that the sum of the first column in $N$ and the third column in $M$ gives the coefficients of the antisymmetric Bianchi identity
\begin{align}
\tilde C_2(\bar D  \wedge\tilde\Theta^-)-3\tilde C_1(\bar D  \wedge\tilde\Theta^{(0)})+2(\tilde C_0-2)(\bar D  \wedge\tilde\Theta^+)=0\;,
\end{align}
while the difference gives those of the symmetric Bianchi identity
\begin{align}
\tilde C_1(\bar D  \vee\tilde\Theta^{(0)})-\tilde C_0(\bar D  \vee\tilde\Theta^+)=0\;.
\end{align}
This is precisely what is obtained when $\Upsilon^-$ is ``folded'' into $\Upsilon^+$ by eq. \eqref{GLambdaIds},
which means that the first column is added to (for antisymmetric product) and subtracted from (for symmetric product) the third column. 
Then, only terms with the identity matrix remain. A general Ansatz containing also such terms reveals that 
\begin{align}
X=N+M+\left(\begin{matrix}-1&0&-1\\0&4&0\\-1&0&-1\end{matrix}\right)\;.
\label{XSolutionEq}
\end{align}
The equality of the first and third columns in the constant matrix guarantees that only $\Upsilon^+\wedge1$ is generated in the variation.
Then, the resulting element $-2$ at position $-+$ compensates for the right hand side of using eq. 
\eqref{ASCtCommutator} for the terms in the variation represented by the first row of $N$, while the resulting $-2$ at 
position $++$ provides the constant term (as in $\tilde C_0-2$) in the antisymmetric Bianchi identity. The $4$ in the middle compensates both for a term from \eqref{ASCtCommutator} for the second row in $N$ and for the constant in $\tilde C_0+2$ in the Bianchi identity of the second column.

%\Red{What can we say about the question of well-definedness of 
%$\brabra {H^{-1}}C_m$?
%Seen as a matrix, it will contain an infinite sum of the form
%\begin{align}
%\sum_{n\in\ZZ}\eta^{AB}t_{A^\star,n}t_{B,m+n}\;.
%\end{align}
%Or is there a matter of order: first act with $C_m$, then contract with $H^{-1}$?
%}

\subsection{Local invariance of ancillary fields}

Compared to the content in the tensor hierarchy algebra, we have introduced one more ancillary field. Instead of only $\ket{\gamma^-}$, we have both $\ket{\tilde\gamma^-}$ and $\ket{\tilde\gamma^+}$, to account for the correct transformations under generalised diffeomorphisms of 
$\ket{\tilde\Theta^\pm}$. We will now show that only the combination $\ket{\tilde\gamma^-+\tilde\gamma^+}$ appears in the action, implying a local symmetry $\delta\ket{\tilde\gamma^-}=\ket\Sigma$, $\delta\ket{\tilde\gamma^+}=-\ket\Sigma$, with $\ket\Sigma$ section-constrained. This symmetry can in principle be used to set $\ket{\tilde\gamma^+}=0$. This implies that one only needs to introduce  a single ancillary vector field $\ket{\gamma^-}$  to define the theory. Such a choice is however inconvenient, as it interferes with the tensorial properties of the torsion under generalised diffeomorphisms.

Consider first the terms in the action quadratic in ancillary fields. All $C_m$'s in the matrix $X$ of eq. 
\eqref{XSolutionEq} give $0$ due to the symmetric section constraint.
The only remainder comes from the $-1$'s in the corner of the last matrix. The contribution to the action is
\begin{align}
-{1\over2}\varrho\brabra{G^{-1}}\ket{\tilde\gamma^-+\tilde\gamma^+}\otimes\ket{\tilde\gamma^-+\tilde\gamma^+}\;.\label{AncAction1}
\end{align}

Next, the contribution linear in ancillary fields is (using the form \eqref{LwithHEq})
\begin{align}
\varrho\brabra{H^{-1}}&\left(
(C_4+C_2-1)(1\otimes t^-_\alpha)-(2C_3+C_1)(1\otimes t_\alpha)+(C_2-1)(1\otimes t^+_\alpha)
\right)\nn\\
&\qquad\times\ket{E^{-1}\tilde\gamma^-}\otimes\ket{E^{-1}\Gamma'^\alpha}\nn\\
+\varrho\brabra{H^{-1}}&\left(
(C_2-1)(1\otimes t^-_\alpha)-2C_1(1\otimes t_\alpha)+(C_0-1)(1\otimes t^+_\alpha)
\right)\label{TauGammaContent}\\
&\qquad\times\ket{E^{-1}\tilde\gamma^+}\otimes\ket{E^{-1}\Gamma'^\alpha}\;,\nn
\end{align}
where $\ket{\Gamma'^\alpha}$ is defined after eq. \eqref{CDThetaEq}.
%Here, we have removed all tilde's for clearer expressions, the calculation is the same.
We want to use the section constraint with one $\ket{E^{-1}\tilde\gamma^\pm}$ and one $\ket{E^{-1}\Gamma'^\alpha}$, and need to pull the $C_m$'s past the operators $1\otimes t_\alpha^{(n)}$. All commutators with $1\vee t^{(n)}_\alpha$ produce $C_n$, $n\geq1$, which yield $0$ due to the (symmetric and antisymmetric) section constraint. The commutators with $1\wedge t^{(n)}_\alpha$ cancel in the third line of eq. \eqref{TauGammaContent}, but in the first line there is a remaining
$[C_2,1\wedge t^-_\alpha]-[C_1,1\wedge t_\alpha]=2(1\wedge t^+_\alpha)$.
The section constraint from the $C_m$ factors, after having passed to the right of $1\otimes t_\alpha^{(n)}$, on the other hand, gives $0$ in the first line, but in the third line $C_0$ gives $1-\varsigma$ for the antisymmetric part. Taken together, the terms linear in ancillary fields are 
%\Red{MC: The signs seem to be correct. The terms without $C_m$'s give $-(1\otimes t^-_\alpha)-(1\otimes t^+_\alpha)$ for both terms in eq. \eqref{TauGammaContent}, then each of them get an additional $2(1\wedge t^+_\alpha)$ by the two different mechanisms described above.}{{\bf G.B: I agree with your text, but $-(1\otimes t^-_\alpha)-(1\otimes t^+_\alpha)+ 2(1\wedge t^+_\alpha) = -(1\otimes t^-_\alpha)-(t^+_\alpha\otimes1 )$ .}} 
\begin{align}
&-\varrho\brabra{H^{-1}}\left((1\otimes t^-_\alpha)+(t^+_\alpha\otimes1)\right)
\ket{E^{-1}(\tilde\gamma^-+\tilde\gamma^+)}\otimes\ket{E^{-1}\Gamma'^\alpha}\nn\\
=&-\varrho\brabra{G^{-1}}\left((1\otimes \tilde t^-_\alpha)+(\tilde t^+_\alpha\otimes1)\right)
\ket{\tilde\gamma^-+\tilde\gamma^+}\otimes\ket{\Gamma^\alpha}\;,\label{AncAction2}
\end{align}
which proves the statement. Note however that the expressions \eqref{AncAction1}, \eqref{AncAction2} are inconvenient to use (except for deriving equations of motion), since they do not maintain manifest symmetry under generalised diffeomorphisms.

\subsection{From coset dynamics to teleparallel dynamics}

The coset dynamics was first formulated in \cite{Bossard:2018utw} for $\gG^+$. It is defined in terms of the coset Maurer-Cartan form 
\begin{align} P_M =  \tfrac12  E^{-1} \partial_M E - \tfrac12 \tau_H \bigl(  E^{-1} \partial_M E \bigr)  - \varrho^{-1} \partial_M \varrho  = - \tfrac12 \Gamma_M{}^\alpha E^{-1} \bigl( t_\alpha   - \tau( t_\alpha  ) \bigr) E   \; , \label{PGamma}\end{align}
and its mode shifted version 
\begin{align} P^{(1)}_M = s P_M + {\chi}_M \; ,\end{align}
that depends on the ancillary field $\chi_M$ satisfying the same constraints as $\tilde{\gamma}^{\pm}_M$. 

We will now demonstrate the equivalence of that formulation and the one in the present paper, 
by showing that the Lagrangians differ by a total derivative.

 The Lagrangian \cite{Bossard:2018utw,Bossard:2023wgg} can be written in our conventions \footnote{The Lagrangian as written in \cite{Bossard:2018utw} differs by a redefinition of $|\chi\rangle$ with the one written here that uses the conventions of \cite{Bossard:2023wgg}. } 
%%%%%%%%%
\begin{multline} \mathscr{L}_{\rm C} =- \varrho \brabra{H^{-1}} \Bigl[ \bigl( \eta_{\alpha\beta}  +2 \delta_\beta^0 \otimes t_\alpha + 4\delta_\beta^1 \otimes s t_\alpha  -2 t_\beta \otimes t_\alpha \bigr)E^{-1} |P^\alpha\rangle \otimes E^{-1} |P^\beta\rangle  \\
+  2 t_\beta \otimes t_\alpha E^{-1} |P^{(1) \alpha} \rangle \otimes E^{-1} |P^{(1) \beta}\rangle \Bigr]\,,  \label{LCLagrange} \end{multline}
%%%%%%%%%
where $\eta_{\alpha\beta}$ is the inverse of $\eta^{(0) \alpha\beta}$ on $\mathfrak{gl}_1 \inplus\fg^+$ and zero for $\alpha =1$ (corresponding to $L_1$). Here we use that $ P_M = P_{M}{}^\alpha t_\alpha $. Recall that this is not true for $\Gamma_M = \Gamma_M{}^\alpha t_\alpha - \Gamma_M{}^0$. We first want to relate $|\chi\rangle$ to $\ket {\gamma^\pm}$. For this we use the property that $s$ commutes with $\gG^+$ up to a central element so that $\ket \chi$ exists such that 
\begin{align} P^{(1)}_M = -  \tfrac12 \Gamma_M{}^\alpha  E^{-1} \bigl(   \tilde{t}^+_\alpha - \tau( \tilde{t}^-_\alpha)  \bigr)   E - \tfrac12 ( \tilde{\gamma}^+_M + \tilde{\gamma}^-_M )\; . \label{P1Gamma} \end{align}
We will now relate the coset formulation of  \cite{Bossard:2018utw} to the teleparallel formulation described in this paper.  We introduce the notation $G_{\alpha\beta}$ for the generalised metric in the adjoint representation, that is defined such that 
\begin{align}  G_{\alpha\beta} \Gamma_\tau^\beta = - \eta_{\alpha\beta} \Gamma^\beta \; , \qquad  \Gamma_\tau^\alpha \tau(t_\alpha) = \Gamma^\alpha t_\alpha \;   \end{align}
for $\Gamma_\tau^\alpha $ the Chevalley conjugate to any $\Gamma^\alpha$. 
Substituting \eqref{PGamma} and \eqref{P1Gamma}  in the Lagrangian \eqref{LCLagrange} one gets 
\begin{multline} \mathscr{L}_{\rm C} =- \frac{\varrho}{2}  \brabra{G^{-1}} \Bigl[ \bigl( \eta_{\alpha\beta} + G_{\alpha\beta} + 2\delta^0_\beta \otimes t_\alpha + 2t_\alpha \otimes \delta^0_\beta + 2 \delta^1_\beta \otimes \tilde{t}^{+}_{\alpha} + 2 \tilde{t}^{-}_{\alpha}  \otimes \delta_\beta^1  \\ - t_\beta \otimes t_\alpha +1  \otimes t_\beta \tau(t_\alpha)  - 2 (1\otimes t_\beta t_\alpha)  \\
  + \tilde{t}^+_\beta \otimes \tilde{t}^+_\alpha -1 \otimes  \tilde{t}^-_\beta \tau(\tilde{t}^-_\alpha)  + 2 (1 \otimes \tilde{t}^-_\beta \tilde{t}^+_\alpha )  \bigr)  |\Gamma^\alpha\rangle \otimes  |\Gamma^\beta\rangle  \\
+ 2  \left(1\otimes \tilde t^-_\alpha+ \tilde t^+_\alpha\otimes1 \right) \ket{\tilde\gamma^-{+}\tilde\gamma^+}\otimes\ket{\Gamma^\alpha} + \ket{\tilde\gamma^-{+}\tilde\gamma^+}\otimes\ket{\tilde\gamma^-{+}\tilde\gamma^+}  \Bigr]\,.  \end{multline}
Using the algebraic identity \cite[Eq. (A.1)]{Bossard:2023wgg}, one then computes that 
\begin{align} & \quad - \frac{\varrho}{2}  \brabra{G^{-1}} \Bigl[ \bigl( G_{\alpha\beta} + 2t_\alpha \otimes \delta^0_\beta + 2 \tilde{t}^{-}_{\alpha}  \otimes \delta_\beta^1  +1 \otimes t_\beta \tau(t_\alpha) - 1 \otimes \tilde{t}^-_\beta \tau(\tilde{t}^-_\alpha)    \bigr)  |\Gamma^\alpha\rangle \otimes  |\Gamma^\beta\rangle  \nonumber \\
& \hspace{30mm}+ 2  \left(1\otimes \tilde t^-_\alpha \right) \ket{\tilde\gamma^-}\otimes\ket{\Gamma^\alpha} + \ket{\tilde\gamma^-}\otimes\ket{\tilde\gamma^-}  \Bigr]\nonumber \\
&= - \frac{\varrho}{2}  \brabra{G^{-1}}  \;  \ket {\tilde{\Theta}^{-}}  \otimes \ket {\tilde{\Theta}^{-}}    + \frac{\varrho}{2}  \brabra{G^{-1}}    \; t_\alpha \ket{\Gamma^\alpha} \otimes t_\beta  \ket{\Gamma^\beta}  \,,  \end{align}
and
\begin{align} &\quad - \frac{\varrho}{2}  \brabra{G^{-1}} \Bigl[ \bigl( \eta_{\alpha\beta} + 2t_\alpha \otimes \delta^0_\beta + 2 \delta^1_\beta \otimes \tilde{t}^{+}_{\alpha}   -  2( 1\otimes t_\beta t_\alpha)   +2 ( 1\otimes \tilde{t}^-_\beta \tilde{t}^+_\alpha )  \bigr)  |\Gamma^\alpha\rangle \otimes  |\Gamma^\beta\rangle  \nonumber \\
& \hspace{30mm} + 2  \left( \tilde t^+_\alpha\otimes1 \right) \ket{\tilde\gamma^-}\otimes\ket{\Gamma^\alpha} + 2  \left( 1\otimes \tilde t^-_\alpha \right) \ket{\tilde\gamma^+}\otimes\ket{\Gamma^\alpha} + 2 \ket{\tilde\gamma^-}\otimes\ket{\tilde\gamma^+}  \Bigr] \nonumber \\
&=  \frac{\varrho}{2}  \brabra{G^{-1}} \Bigl[  \tilde{C}_2  \ket {\tilde{\Theta}^{-}}  \otimes  \ket {\tilde{\Theta}^{-}}   -2 \tilde{C}_1 \ket {\theta}  \otimes \ket {\tilde{\Theta}^{-}} + \tilde{C}_0  \ket {\theta} \otimes \ket {\theta}  - 2 \ket {\tilde{\Theta}^{+}}  \otimes \ket {\tilde{\Theta}^{-}}    \Bigr] \nonumber \\ 
& \qquad -  \frac{\varrho}{2}  \brabra{G^{-1}} ( t_\alpha \otimes t_\beta + t_\beta \otimes t_\alpha ) \ket \Gamma^\alpha \otimes \ket \Gamma^\beta  \, .  \end{align}
Using finally 
\begin{align} &\quad - \frac{\varrho}{2}  \brabra{G^{-1}} \Bigl[ \bigl(  \tilde{t}^+_\beta \otimes \tilde{t}^+_\alpha  \bigr)  |\Gamma^\alpha\rangle \otimes  |\Gamma^\beta\rangle  + 2  \left( \tilde t^+_\alpha\otimes1 \right) \ket{\tilde\gamma^+}\otimes\ket{\Gamma^\alpha} +  \ket{\tilde\gamma^+}\otimes\ket{\tilde\gamma^+}  \Bigr]  \\
&=  \frac{\varrho}{2}  \brabra{G^{-1}} \Bigl[  \tilde{C}_4  \ket {\tilde{\Theta}^{-}}  \otimes  \ket {\tilde{\Theta}^{-}}   - 4 \tilde{C}_3 \ket {\theta}  \otimes \ket {\tilde{\Theta}^{-}} +4 \tilde{C}_2  \ket {\theta} \otimes \ket {\theta} \nonumber \\ 
& \hspace{20mm} + 2 \tilde{C}_2 \ket{\tilde{\Theta}^{-}} \otimes  \ket{\tilde{\Theta}^{+}}- 4 \tilde{C}_1  \ket {\theta} \otimes \ket{\tilde{\Theta}^{+}} + \tilde{C}_0  \ket{\tilde{\Theta}^{+}} \otimes \ket{\tilde{\Theta}^{+}}  - \ket {\tilde{\Theta}^{+}}  \otimes \ket {\tilde{\Theta}^{+}}      \Bigr] \,,  \nonumber \end{align}
one obtains that $\mathscr{L}_{\rm C} $ coincides with \eqref{XLagrangian} up to a total derivative 
\begin{align} \mathscr{L}_{\rm C}  =  \mathscr{L}  + \frac{\varrho}{2} \brabra{G^{-1}}  ( 1\otimes t^\alpha + t^\alpha \otimes 1) |\partial\rangle \otimes |\Gamma_\alpha  \rangle\,, \end{align}
where $|\partial\rangle$ acts on all terms. 

Note that the expression of the Lagrangian in \cite{Bossard:2018utw} involves naively infinitely many terms associated to the infinitely many components of $|\Gamma^\alpha\rangle $ in the adjoint representation. In the teleparallel formulation it is manifest that the Lagrangian only involves finitely many components of $|\Gamma^\alpha\rangle $ for any solution to the section constraint. This was checked explicitly in \cite{Bossard:2018utw}. However, the exceptional field theory Lagrangian describing the entire dynamics in $2+9$ or $2+8$ dimensions does involve all the components of the coset fields and cannot be written as a finite expression with manifest $\gG^+$ symmetry \cite{Bossard:2023wgg}. 

%Let us finally comment on the fact that the teleparallel Lagrangian is especially well suited for generalised Scherk--Schwarz reduction \cite{Hohm:2014qga}. In %particular, the form of the supergravity potential derived in \cite{Bossard:2023wgg} follows directly in the teleparallel formulation. 

%\subsection{Physical degrees of freedom\label{PhysicalSection}}

\section{Outlook\label{OutlookSection}}

One main motivation for the present work is to identify features of extended geometry that are inherent to infinite-dimensional structure groups, and may be used as guidelines when searching for the appropriate formulation with \eg\ hyperbolic groups, corresponding to BKL symmetry, or even very extended groups, such as $E_{11}$ \cite{West:2001as,Bossard:2017wxl}.
One such feature is the presence of more generators than contained in the naive structure algebra. This is clear already from inspection of the relevant tensor hierarchy algebra. In the present case, this is the generator $L_1$. 
Another interesting feature that distinguishes the present case from finite-dimensional ones is the r\^ole of symmetric Bianchi identities for the demonstration of local invariance. The possibility arises to form elements in the local compact subalgebra from the symmetric product of fundamentals, not only antisymmetric.
This will certainly persist in any further extension of the structure algebra.
 
One phenomenon that makes predictions for further extended structure algebras more difficult is the slight mismatch between the torsion predicted by the tensor hierarchy algebra $S(\fg^{++})$ and the one we have used. In particular, only $\Theta^-$ and $\theta$ are found in the tensor hierarchy algebra, while we introduced also $\Theta^{(n)}$, $n>0$. In the end, torsion was packaged into $\tilde\Theta^-$, $\theta$ and $\tilde\Theta^+$, which still accounts for one more fundamental than predicted by the tensor hierarchy algebra.
There are a couple of possible interpretations. One would be that $\tilde\Theta^+$ in some sense is ``auxiliary'', and that it is possible to formulate the dynamics without it. We have tried, but have not been able to find such a formulation, for the simple reason that the opportunity to relate the transformations of $\tilde\Theta^-$ and $\tilde\Theta^+$ using $\tau_H(\Upsilon)=\Upsilon$ then is lost. Another possibility is that the affine algebra is a particular, in some sense singular, case, having some properties that disappear when further extending. Indeed, the Cartan matrix of an affine Kac--Moody algebra is singular, as is that of the tensor hierarchy algebra extension $S(\fg^{++})$ of an over-extended Kac--Moody algebra.
There is also the possibility that the appropriate tensor hierarchy algebra is larger than what we have anticipated, but we are not aware of such an algebra.
In any case, we need to proceed with a certain awareness that this kind of surprises may extend to situations with further extended 
structure algebras.
 
We have learnt that a useful way of forming invariants of the full extended structure algebra (\ie, including $L_1$), and to form tensors in a way that unmakes the indecomposable, Jordan cell, type of transformations of the linear fields, is to use explicit conjugation with the vielbein. This may well continue to be a valuable tool.

When investigating the level decomposition of a tensor hierarchy algebra $S(\fg^{+++})$, relevant to BKL geometry, one finds that the over-extended 
Kac--Moody algebra $\fg^{++}$ is complemented with a lowest weight fundamental $V$
\cite{Cederwall:2021ymp}. $L_1$ can be seen as its lowest weight state.
The resulting extended structure algebra is $\fg^{++}\oplus V$ as a vector space, but is not a semi-direct sum as a Lie algebra.
Another new feature is the presence of extra generators complementing the generalised diffeomorphisms in $V$ with a subleading symmetric module. An interesting step towards extended geometry with a very extended structure algebra has been taken in ref. \cite{Bossard:2021ebg}. However, it lacks some of these ingredients. We expect the full structure to be relevant, for example for an algebraic understanding of the emergence of space (or space-time) in terms of gradient structures 
\cite{Kleinschmidt:2005bq} in the algebra.

\vspace{2\parskip}\noindent\underbar{Acknowledgements:} MC and JP would like to thank Benedikt K\"onig, Axel Kleinschmidt, Hermann Nicolai and Alex J. Feingold for discussions on branching of $\fg^+$ modules under the $\fk(\fg^+)$ subalgebra. GB would like to thank Franz Ciceri, Gianluca Inverso and Axel Kleinschmidt for useful discussions. 

%\begin{appendices}

%\newpage

\appendix

\section{Tensor hierarchy algebra extensions of over-extended Kac--Moody algebras\label{THAApp}}

Tensor hierarchy algebras are constructed and investigated in refs. 
\cite{Palmkvist:2013vya,Bossard:2017wxl,Carbone:2018xqq,Cederwall:2019qnw,Cederwall:2021ymp,Cederwall:2022oyb,Cederwall:2023stz}, and their r\^ole in extended geometry developed in refs. \cite{Palmkvist:2013vya,Cederwall:2018aab,Cederwall:2019bai,Cederwall:2023xbj}.
%Let the (untwisted) affine algebra be $\fg^+$, for a finite-dimensional Lie algebra $\fg$. The underlying superalgebra, encoding gauge parameters, fields, torsion etc. is then the tensor hierarchy algebra 
%$S(\fg^{++})$. This superalgebra was described in ref. \cite{Cederwall:2021ymp}, and we use its double grading with respect to the fermionic and over-extended nodes, where each level contains a module of the subalgebra at level $(0,0)$. This subalgebra is the centrally extended loop algebra of $\fg$, complemented with the Cartan generator $\dd$ (``$L_0$''), and further extended by $L_1$.
We will in this appendix look closer at the tensor hierarchy algebra $S(\fg^{++})$
in the double grading with degrees $(p,q)$ described in Section~\ref{FieldsFromTHASec}. The results in the
appendix
thus extends Section~7.4 of ref. \cite{Cederwall:2021ymp}. We refer to this paper for other gradings and aspects of
the tensor hierarchy algebra $S(\fg^{++})$.

Modules of $\fg^{++}$ are found at SW-NE diagonals (with constant $p-q$). The diagonal $p-q=0$ consists of the extension of $\fg^{++}$, with basis elements
\begin{align}
T_{\underline\alpha}=(\ldots,\bar A'^{\flat MN},F^{\flat M},(T_{A,m},\dd,\KK),E^\sharp_M,
A^\sh_{MN},\ldots)\;,
\end{align}
extended by its lowest weight fundamental, with basis elements
\begin{align}
J_{\underline M}=(L_1,L_M,(S'^\sh_{MN},A'_{MN}),\ldots)\:.
\end{align}
%We use the letters $A$ and $S$ to indicate that the elements transform in the subleading submodules 
%of the antisymmetric and symmetric tensor product, respectively. They are obtained by factoring out the leading
%submodules
%$R(\pm2\lambda\mp\alpha_0)$ and $R(\pm2\lambda)$ from the full
%antisymmetric and symmetric tensor product, respectively.

\begin{table}
  \begin{align*}
  \xymatrix@=.3cm{
    \ar@{-}[]+<1.4em,1em>;[dddddd]+<1.4em,-1em>
    \ar@{-}[]+<-0.8cm,-1em>;[rrrrr]+<1.4cm,-1em>
   q  &\ar@{-}[]+<4.8em,1em>;[dddddd]+<4.8em,-1em> p=-2
     &\ar@{-}[]+<2.65em,1em>;[dddddd]+<2.65em,-1em> p=-1
    & \ar@{-}[]+<3.5em,1em>;[dddddd]+<3.5em,-1em> p=0 
    & \ar@{-}[]+<2.5em,1em>;[dddddd]+<2.5em,-1em> p=1 &p=2\\
3&&&&&\shift{4}{A'^\sh_{MN}}\\
2&&&\shift{1}{\pi^\sh}&\shift{2}{L^\sh_M}
&\shift{4}{A^\sh_{MN}}\;\shift{3}{S'^\sh_{MN}}\;\shift{3}{A'_{MN}}
\\
1&\shift{-2}{\bar S^{MN}}&\shift{-1}{\Phi^{\sh M}}&\shift{1}{\KK^\sh}\;
       \shift{0}{\pi}\;\shift{0}{L_1^\sh}\;{T^{\sh A}_m}
   & \shift{2}{E_M^\sh}\;\shift{1}{L_M}
   &\shift{3}{S^\sh_{MN}}\;\shift{3}{A_{MN}}\;\shift{2}{S'_{MN}}
       \\ 
0&\shift{-3}{\bar S^{\flat MN}}\;\shift{-3}{\bar A^{MN}}\;\shift{-2}{\bar S'^{MN}}&\shift{-2}{\Phi^M}\;\shift{-1}{F^M}
&\shift{0}{\KK}\;\shift{0}{\dd}\;\shift{-1}{L_1}\;T_{Am}
           & \shift{1}{E_M}&\shift{2}{S_{MN}}\\ 
-1 &\shift{-4}{\bar A^{\flat MN}}\;\shift{-3}{\bar S'^{\flat MN}}\;\shift{-3}{\bar A'^{MN}}& \shift{-2}{F^{\fl M}}&
\shift{-1}{\epsilon_{-2}} &&\\
-2&\shift{-4}{\bar A'^{\flat MN}}&&&&
}
\end{align*}
  \caption{\it Basis elements for $S(\fg^{++})$ for $-2\leq p\leq2$. The weights specifying the action of $\dd$ are given in red.}
\label{HyperbolicSTableBasis}
\end{table}

The index $M$ used for all fundamental or anti-fundamental modules
is covariant under the centrally extended loop algebra, \ie,
\begin{align}
  [T_{A,m},X_M]&=-(t_{A,m})_M{}^N X_N\;,
     &[T_{A,m},Y^M]&=(t_{A,m})_N{}^MY^N\;,\nn\\
  [\KK,X_M]&=-X_M\;,&[\KK,Y^M]&=Y^M\;.
\end{align}
The action of $\dd$ depends on the weight/mode number shift:
\begin{align}
[\dd,X_M]&=-(\ell_0)_M{}^NX_N+w(X)X_M\;,\nn\\
[\dd,Y^M]&=(\ell_0)_N{}^MY^N+w(Y)Y^M\;.\label{dWeightTransf}
\end{align}
Note that $\dd^\flat=-[\dd,\epsilon_{-2}]=\epsilon_{-2}$, so $w(\epsilon_{-2})=-1$,
and $w(X^\flat)=w(X)-1$. There is an arbitrariness in the assignment of weight due to redefinitions $\dd\mapsto \dd+a\KK$. The convention used here corresponds to canonical (tensorial) weights in the extended geometry.

The modules at $p=2$ with basis elements denoted $S$ and $A$ consist of all modules in the symmetric and antisymmetric tensor products of the lowest weight fundamental except the leading one. 

From the over-extended Kac--Moody algebra on the diagonal, we have
\begin{align}
  [E^\sh_M,F^{\fl N}]&=\delta_M{}^N\dd+(\ell_0)_M{}^N\KK
  -\sum_{m\in\ZZ}\eta^{AB}(t_{A,m})_M{}^NT_{B,-m}\nn\\
  &=-\eta^{(0)\alpha\beta}t_{\alpha M}{}^NT_\beta\;.
  \label{EshFflBracket}
\end{align}
%deriving from the combination
%$\Cn0=\KK\otimes\dd+\dd\otimes\KK-\sum_{m\in\ZZ}\eta_{AB}T^A_m\otimes T^B_{-m}$.
Due to the non-invariance of $\eta^{(0)}$ under $L_1$,  
the bracket \eqref{EshFflBracket} is {\it not} consistent
on its own, unless the action of $L_1$ is modified. The Jacobi identity is not consistent with $L_1$ acting as\footnote{We use ``$\bullet$'' to indicate an equation that does not hold.}
$\bullet[L_1,E^\sh_M]=-(\ell_1)_M{}^NE^\sh_N$ and
$[L_1,F^{\fl M}]=(\ell_1)_N{}^MF^{\fl N}$. Then one would get
\begin{align}
  &\bullet[L_1,[E^\sh_M,F^{\fl N}]]-[[L_1,E^\sh_M],F^{\fl N}]
  -[E^\sh_M,[L_1,F^{\fl N}]]\nn\\
  &=\delta_M^NL_1+(\ell_1)_M{}^N\KK-\sum_{m\in\ZZ}\eta^{AB}(t_{A,m})_M{}^NT_{B,1-m}\\
  &=-\eta^{(1)\alpha\beta}t_{\alpha M}{}^NT_\beta\;.\nn
\end{align}
%We recognise the $\Cn1$ structure.
The remedy is to let 
\begin{align}
  [L_1,E^\sh_M]=-(\ell_1)_M{}^NE^\sh_N-L_M\;,
\end{align}
and
\begin{align}
[L_M,F^{\fl N}]=
\eta^{(1)\alpha\beta}t_{\alpha M}{}^NT_\beta\;.
%=-\delta_M^NL_1-(\ell_1)_M{}^N\KK+\sum_{m\in\ZZ}(t^A_m)_M{}^NT^A_{1-m}\;.
\end{align}
The action of $\dd$ and $L_1$ on this last commutator fulfil
Jacobi identities.
In the same way, one obtains
\begin{align}
[L_1,F^M]=(\ell_1)_N{}^MF^N-\Phi^M\label{JordanFPhi}
\end{align}
and
\begin{align}
  [E_M,\Phi^N]=
  \eta^{(1)\alpha\beta}t_{\alpha M}{}^NT_\beta\;.
  %-\delta_M^NL_1-(\ell_1)_M{}^N\KK+\sum_{m\in\ZZ}(t^A_m)_M{}^NT^A_{1-m}\;.
\end{align}

We see that $L_1$, which is the lowest state in the ``extra'' module appearing
together with the adjoint of $\fg^{++}$ on the diagonal, transforms the
states $E^\sh_M$ in the adjoint both to themselves and to $L_M$, which
is in the ``extra'' module.
At a given $(p,q)$, we have a module of the $(0,0)$ subalgebra, which
in general is not completely reducible, but has a Jordan cell
structure. For example, at $(p,q)=(1,1)$,
\begin{align}
  \left[L_1,\left(\begin{matrix}E^\sh_M\\L_M\end{matrix}\right)\right]
  =-\left(\begin{matrix}(\ell_1)_M{}^N&\delta_M{}^N\\0
    &(\ell_1)_M{}^N\end{matrix}\right)
    \left(\begin{matrix}E^\sh_N\\L_N\end{matrix}\right)\;,
\end{align}
and at $(p,q)=(0,1)$,
\begin{align}
  \left[L_1,\left(\begin{matrix}\KK^\sh\\ \pi\end{matrix}\right)\right]
  =\left(\begin{matrix}0&1\\0&0\end{matrix}\right)
    \left(\begin{matrix}\KK^\sh\\ \pi\end{matrix}\right)\;.
\end{align}
All this is of course precisely what is obtained from ref. \cite{Cederwall:2021ymp}. %Could leave out a lot of it.

Brackets between $p=0$ and $p=\pm1$, beyond the ones that are covariant, or vanish
by consistency of the weights:
\begin{align}
  [\KK^\sh,E_M]&=E^\sh_M\;,&[\KK^\sh,F^{\fl M}]&=F^M\;,\nn\\
  [\KK^\sh,L_M]&=-L^\sh_M\;,&[\KK^\sh,\Phi^M]&=\Phi^{\sh M}\;,\nn\\
  [T^\sh_{A,m},E_M]&=(t_{A,m})_M{}^NE^\sh_N\;,  %+(t_{A,m-1})_M{}^NL_N
  &[T^\sh_{A,m},F^{\fl M}]&=(t_{A,m})_N{}^MF^N\nn\\
  &\quad\,+(t_{A,m-1})_M{}^NL_N\;, & &\quad\,+(t_{A,m-1})_N{}^M\Phi^N\;,\nn\\
  [T^\sh_{A,m},E^\sh_M]&=(t_{A,m-1})_M{}^NL^\sh_N\;,
  &[T^\sh_{A,m},F^M]&=(t_{A,m-1})_N{}^M\Phi^{\sh N}\nn\;,\\
  [T^\sh_{A,m},L_M]&=-(t_{A,m})_M{}^NL^\sh_N\;,
       &[T^\sh_{A,m},\Phi^M]&=-(t_{A,m})_N{}^M\Phi^{\sh N}\;,\\
  [L_1^\sh,E_M]&=(\ell_1)_M{}^NE^\sh_N+(\ell_0+1)_M{}^NL_N\;,
      &[L_1^\sh,F^{\fl M}]&=(\ell_1)_N{}^MF^N+(\ell_0)_N{}^M\Phi^N\nn\;,\\
  [L_1^\sh,E^\sh_M]&=(\ell_0)_M{}^NL^\sh_N\;,
        &[L_1^\sh,F^M]&=(\ell_0+1)_N{}^M\Phi^{\sh N} \nn\;,\\
  [L_1^\sh,L_M]&=-(\ell_1)_M{}^NL^\sh_N\;,
       &[L_1^\sh,\Phi_M]&=-(\ell_1)_N{}^M\Phi^{\sh N}\nn\;,\\ 
  [\pi,E_M]&=-L_M\;,   &[\pi,F^{\fl M}]&=\Phi^M\nn\;,\\
  [\pi,E^\sh_M]&=-L^\sh_M\;,   &[\pi,F^M]&=\Phi^{\sh M}\nn\;,\\
  [\pi^\sh,E_M]&=L^\sh_M\;,   &[\pi^\sh,F^{\fl M}]&=\Phi^{\sh M}\nn\;.
\end{align}
Non-obvious brackets between generators at $p=0$:
\begin{align}
  [\KK^\sh,\pi]&=0\;,\nn\\
  [\KK^\sh,L_1^\sh]&=-\pi^\sh\;,\nn\\
  [T^A_m,T^{\sh B}_n]&=f^{AB}{}_CT^{\sh C}_{m+n}
  +\eta^{AB}m\delta_{m+n,0}\KK^\sh+\eta^{AB}m\delta_{m+n-1,0}\pi\;,\\
  [T^{\sh A}_m,T^{\sh B}_n]&=-\eta^{AB}\delta_{m+n-1,0}\pi^\sh\;.\nn
\end{align}
%[Comment on the possibility of adding $\pi$ to $L_1^\sh$.]

The brackets between
generators at $p=1$ and $p=-1$ are all based on the $\eta^{(0)}$
and $\eta^{(1)}$ structures:
\begin{align}
  [E_M,F^{\fl N}]&=\delta_M^Ne_0\;,\nn\\
  [E_M,F^N]&=\delta_M^N\KK-\eta^{(0)\alpha\beta}t_{\alpha M}{}^NT_\beta\;,\nn\\
  %\delta_M{}^N\dd+(\ell_0+1)_M{}^N\KK -\sum_{m\in\ZZ}\eta_{AB}(t^A_m)_M{}^NT^B_{-m}\;,\\
  [E^\sh_M,F^{\fl N}]&=
  %\delta_M{}^N\dd+(\ell_0)_M{}^N\KK-\sum_{m\in\ZZ}\eta_{AB}(t^A_m)_M{}^NT^B_{-m}
  -\eta^{(0)\alpha\beta}t_{\alpha M}{}^NT_\beta\;,\nn\\
  [E^\sh_M,F^N]&=-\delta_M^N\KK^\sh\;,\nn\\
  [E_M,\Phi^N]&=\eta^{(1)\alpha\beta}t_{\alpha M}{}^NT_\beta\;,\nn\\
  %-\delta_M^NL_1-(\ell_1)_M{}^N\KK+\sum_{m\in\ZZ}(t^A_m)_M{}^NT^A_{1-m}\;,\\
  [E_M,\Phi^{\sh N}]
  &=\eta^{(1)\alpha\beta}t_{\alpha M}{}^NT^\sh_\beta
  %-\delta_M^NL_1^\sh-(\ell_1)_M{}^N\KK^\sh+\sum_{m\in\ZZ}(t^A_m)_M{}^NT^{\sh A}_{1-m}
  -(\ell_0+1)_M{}^N\pi\;,\nn\\
  [E^\sh_M,\Phi^N]&=
  %\delta_M^NL_1^\sh+(\ell_1)_M{}^N\KK^\sh -\sum_{m\in\ZZ}(t^A_m)_M{}^NT^{\sh A}_{1-m}
  -\eta^{(1)\alpha\beta}t_{\alpha M}{}^NT^\sh_\beta
      -(\ell_0)_M{}^N\pi\;,\\
  [E^\sh_M,\Phi^{\sh N}]&=-\delta_M^N\pi^\sh\;,\nn\\
  [L_M,F^{\fl N}]&=\eta^{(1)\alpha\beta}t_{\alpha M}{}^NT_\beta\;,\nn\\
  %-\delta_M^NL_1-(\ell_1)_M{}^N\KK+\sum_{m\in\ZZ}(t^A_m)_M{}^NT^A_{1-m}\;,\\
  [L_M,F^N]
  &=\eta^{(1)\alpha\beta}t_{\alpha M}{}^NT^\sh_\beta
  %-\delta_M^NL_1^\sh-(\ell_1)_M{}^N\KK^\sh+\sum_{m\in\ZZ}(t^A_m)_M{}^NT^{\sh A}_{1-m}
 +(\ell_0+1)_M{}^N\pi\;,\nn\\
  [L^\sh_M,F^{\fl N}]&=\eta^{(1)\alpha\beta}t_{\alpha M}{}^NT^\sh_\beta
  % -\delta_M^NL_1^\sh-(\ell_1)_M{}^N\KK^\sh+\sum_{m\in\ZZ}(t^A_m)_M{}^NT^{\sh A}_{1-m}
      +(\ell_0)_M{}^N\pi\;,\nn\\
  [L^\sh_M,F^N]&=\delta_M^N\pi^\sh\;,\nn\\
  [L_M,\Phi^N]&=[L_M,\Phi^{\sh N}]
        =[L^\sh_M,\Phi^N]=[L^\sh_M,\Phi^{\sh N}]=0\;.\nn
\end{align}

\section{Proof of Bianchi identities\label{BianchiProofApp}}

%\Red{Give one example. Fix notation. State which BI's require the M--C equation.}
Of the Bianchi identities among the ones we have stated, only the first one in eq. \eqref{BIEq} and eq. \eqref{UnnecBI} rely on the Maurer--Cartan equation \eqref{MCeq} for $\Gamma$. All rely on the section constraint.
The proofs are all analogous, so we take one example.
The antisymmetric Bianchi identity we want to display the proof of is
\begin{align}
(\cn0-2)\ket{D+\theta}\wedge\ket{\Theta^-}-\cn{-1}\ket D\wedge\ket\theta=0\;.
\end{align}
%It is easily shown that all terms with $\*\vee\Gamma^\alpha$ vanish using the section constraint. 
We check the Bianchi identity directly, by using the Maurer--Cartan equation for terms 
$\ket\*\wedge\ket{\Gamma^\alpha}=-{1\over2}f_{\beta\gamma}{}^\alpha\ket{\Gamma^\beta}\wedge\ket{\Gamma^\gamma}$ and expanding the covariant derivatives as well as the expressions for $\ket{\Theta^-}$ and $\ket\theta$. The result is a sum of terms quadratic in connections.
The appropriate connection terms containing the weights of the shifted torsion components, as well as the Jordan cell behaviour under $L_1$ are of course needed.
The proof of the identity requires repeated use of eq. \eqref{ASCtCommutator}.

%{\it Below not changed to new conventions, yet. Replace by complete calculation.}

Let the torsion $\ket{\Theta^-}$ and $\ket\theta$ be defined by eqs. (\ref{ThetaDef},\ref{ThetaMinusDef}). Then,
\begin{align}
\ket D\otimes\ket{\Theta^-}&=\ket\*\otimes\ket{\gamma^-}+(1\otimes t_\alpha)\ket{\Gamma^\alpha}\otimes\ket{\gamma^-}-2\ket{\Gamma^0}\otimes\ket{\gamma^-}\nn\\
&\quad+(1\otimes t^-_\alpha)\ket\*\otimes\ket{\Gamma^\alpha}
+(1\otimes t_\alpha t^-_\beta)\ket{\Gamma^\alpha}\otimes\ket{\Gamma^\beta}\nn\\
&\quad-2(1\otimes t^-_\alpha)\ket{\Gamma^0}\otimes\ket{\Gamma^\alpha}
-(1\otimes t_\alpha)\ket{\Gamma^{1}}\otimes\ket{\Gamma^\alpha}\;,\\
\ket\theta\otimes\ket{\Theta^-}&=(t_\alpha\otimes1)\ket{\Gamma^\alpha}\otimes\ket{\gamma^-}
+(t_\alpha\otimes t^-_\beta)\ket{\Gamma^\alpha}\otimes\ket{\Gamma^\beta}\;,\nn\\
\ket D\otimes\ket\theta&=(1\otimes t_\alpha)\ket\*\otimes\ket{\Gamma^\alpha}
+(1\otimes t_\alpha t_\beta)\ket{\Gamma^\alpha}\otimes\ket{\Gamma^\beta}
-(1\otimes t_\alpha)\ket{\Gamma^0}\otimes\ket{\Gamma^\alpha}\;.\nn
\end{align}
An important feature is that certain terms in $\ket{D+\theta}\otimes\ket{\Theta^-}$ combine into symmetrised tensor products:
\begin{align}
\ket{D+\theta}\otimes\ket{\Theta^-}&=\ket\*\otimes\ket{\gamma^-}+2(1\vee t_\alpha)\ket{\Gamma^\alpha}\otimes\ket{\gamma^-}-2\ket{\Gamma^0}\otimes\ket{\gamma^-}\nn\\
&+(1\otimes t^-_\alpha)\ket\*\otimes\ket{\Gamma^\alpha}
+2(1\vee t_\alpha)(1\otimes t^-_\beta)\ket{\Gamma^\alpha}\otimes\ket{\Gamma^\beta}\\
&-2(1\otimes t^-_\alpha)\ket{\Gamma^0}\otimes\ket{\Gamma^\alpha}
-(1\otimes t_\alpha)\ket{\Gamma^{1}}\otimes\ket{\Gamma^\alpha}\;.\nn
\end{align}

We can now check all terms in $(\cn0-2)\ket{D+\theta}\wedge\ket{\Theta^-}-\cn{-1}\ket D\wedge \ket\theta$.
Begin with the ones containing $\ket{\gamma^-}$, which are
\begin{align}
(\cn0-2)\left[\ket\*\wedge\ket{\gamma^-}+2(1\vee t_\alpha)\ket{\Gamma^\alpha}\wedge\ket{\gamma^-}-2\ket{\Gamma^0}\wedge\ket{\gamma^-}\right]\;.
\end{align}
All three terms vanish, the first and last from the antisymmetric section constraint. In the second term, 
$[\cn0,2(1\vee t_\alpha)]=-\delta_\alpha^{1}\cn1$, which also annihilates $\ket{\Gamma^\alpha}\wedge\ket{\gamma^-}$.

Next, the terms with $\ket\*\otimes\ket{\Gamma^\alpha}$ are
\begin{align}
&\left[(\cn0-2)(1\wedge t^-_\alpha)-\cn{-1}(1\wedge t_\alpha)\right]\ket\*\vee\ket{\Gamma^\alpha}\nn\\
&\quad+\left[(\cn0-2)(1\vee t^-_\alpha)-\cn{-1}(1\vee t_\alpha)\right]\ket\*\wedge\ket{\Gamma^\alpha}\;.
\end{align}
The first row gives
$(-[1\wedge t^-_\alpha,\cn0]+[1\wedge t_\alpha,\cn{-1}]-2(1\wedge t^-_\alpha))\*\vee\Gamma^\alpha$
using the symmetric section constraint.
This vanishes thanks to eq. \eqref{ASCtCommutator}.
The first term in the second row vanishes, since the only $t^-_\alpha$ not commuting with $\cn0-2$ is
$t^-_0=\ell_{-1}$, but $\ket\*\wedge\ket{\Gamma^0}=0$.
The last term gives ${1\over2}f_{\beta\gamma}{}^\alpha\cn{-1}(1\vee t_\alpha)\ket{\Gamma^\beta}\wedge\ket{\Gamma^\gamma}$
through the Maurer--Cartan equation.
For the rest of the terms, it is convenient to consider $\ket{\Gamma^\alpha}\vee\ket{\Gamma^\beta}$ and
$\ket{\Gamma^\alpha}\wedge\ket{\Gamma^\beta}$ separately.
The terms containing $\ket{\Gamma^\alpha}\vee\ket{\Gamma^\beta}$ are:
\begin{align}
&(\cn0-2)\left[2(1\vee t_\alpha)(1\wedge t^-_\beta)\ket{\Gamma^\alpha}\vee\ket{\Gamma^\beta}
-2(1\wedge t^-_\alpha)\ket{\Gamma^0}\vee\ket{\Gamma^\alpha}
-(1\wedge t_\alpha)\ket{\Gamma^1}\vee\ket{\Gamma^\alpha}\right]\nn\\
&\quad-\cn{-1}\left[2(1\vee t_\alpha)(1\wedge t_\beta)\ket{\Gamma^\alpha}\vee\ket{\Gamma^\beta}
-(1\wedge t_\alpha)\ket{\Gamma^0}\vee\ket{\Gamma^\alpha}\right]\;.\label{calceq0}
\end{align}
The first term on the first row is rewritten as
\begin{align}
2(1\vee t_\alpha)(\cn0-2)(1\wedge t^-_\beta)\ket{\Gamma^\alpha}\vee\ket{\Gamma^\beta}
-\cn1(1\wedge t^-_\beta)\ket{\Gamma^1}\vee\ket{\Gamma^\beta}\;,\label{calceq1}
\end{align}
and the first term on the second row as
\begin{align}
&-2(1\vee t_\alpha)\cn{-1}(1\wedge t_\beta)\ket{\Gamma^\alpha}\vee\ket{\Gamma^\beta}
+\cn{-1}(1\wedge t_\beta)\ket{\Gamma^0}\vee\ket{\Gamma^\beta}\nn\\
&+2\cn0(1\wedge t_\beta)\ket{\Gamma^1}\vee\ket{\Gamma^\beta}\;.\label{calceq2}
\end{align}
The first terms in eqs. \eqref{calceq1} and \eqref{calceq2} cancel using eq. \eqref{ASCtCommutator} and the symmetric section constraint. The remaining terms from eq. 
\eqref{calceq0} contain at least one $\ket{\Gamma^0}$ or $\ket{\Gamma^1}$:
\begin{align}
&2\left[-\cn0(1\wedge t^-_\alpha)+\cn{-1}(1\wedge t_\alpha)+2(1\wedge t^-_\alpha)\right]\ket{\Gamma^0}\vee\ket{\Gamma^\alpha}\nn\\
&+\left[-\cn1(1\wedge t^-_\alpha)+\cn0(1\wedge t_\alpha)+2(1\wedge t_\alpha)\right]\ket{\Gamma^1}\vee\ket{\Gamma^\alpha}=0\;.
\end{align}

Finally, when we collect the terms with $\ket{\Gamma^\alpha}\wedge\ket{\Gamma^\beta}$, the first connection term in $\ket D\wedge\ket\theta$ immediately cancels the contribution from the Maurer--Cartan equation above. Remaining terms are
\begin{align}
&(\cn0-2)\left[2(1\vee t_\alpha)(1\vee t^-_\beta)\ket{\Gamma^\alpha}\wedge\ket{\Gamma^\beta}
-2(1\vee t^-_\alpha)\ket{\Gamma^0}\wedge\ket{\Gamma^\alpha}-(1\vee t_\alpha)\ket{\Gamma^1}\wedge\ket{\Gamma^\alpha}\right]\nn\\
&+\cn{-1}(1\vee t_\alpha)\ket{\Gamma^0}\wedge\ket{\Gamma^\alpha}\;.
\end{align}
The second and third terms vanish. The first term is rewritten as
\begin{align}
&2(1\vee t_\alpha)(\cn0-2)(1\vee t^-_\beta)\ket{\Gamma^\alpha}\wedge\ket{\Gamma^\beta}
-\cn1(1\vee t^-_\beta)\ket{\Gamma^1}\wedge\ket{\Gamma^\beta}\nn\\
&=(1\vee t_\alpha)\cn{-1}\ket{\Gamma^\alpha}\wedge\ket{\Gamma^0}-\cn0\ket{\Gamma^1}\wedge\ket{\Gamma^0}\;,
\end{align}
so together with the fourth term we get
\begin{align}
[\cn{-1},1\vee t_\alpha]\ket{\Gamma^0}\wedge\ket{\Gamma^\alpha}+\cn0\ket{\Gamma^0}\wedge\ket{\Gamma^1}=0\;.
\end{align}

\newpage

\bibliographystyle{utphysmod2}

%\bibliography{biblio.bib}

%\bibliography{biblio}

\providecommand{\href}[2]{#2}\begingroup\raggedright\endgroup

\end{document}